\documentclass[draftclsnofoot,12pt,onecolumn]{IEEEtran}  

\usepackage{amsmath,amsfonts}
\usepackage{algorithmic}
\usepackage{algorithm}
\usepackage{array}
\usepackage{textcomp}
\usepackage{stfloats}
\usepackage{url}
\usepackage{verbatim}
\usepackage{graphicx}
\usepackage{cite}
\usepackage{subfigure}
\usepackage{epstopdf}
\usepackage[justification=centering]{caption}  
\usepackage{color}
\usepackage{setspace}

\begin{document}

\title{Channel Modeling for UAV-to-Ground Communications with Posture Variation and Fuselage Scattering Effect}

\author{Boyu Hua,~\IEEEmembership{Member,~IEEE,} Haoran Ni, Qiuming Zhu{*},~\IEEEmembership{Member,~IEEE,} Cheng-Xiang Wang{*},~\IEEEmembership{Fellow,~IEEE,}
Tongtong Zhou, Kai Mao,~\IEEEmembership{Student Member,~IEEE,} Junwei Bao, Xiaofei Zhang,~\IEEEmembership{Member,~IEEE}
\thanks{This work was supported by National Natural Science Foundation of China, No. 62271250, National Key Scientific Instrument and Equipment Development Project, No. 61827801, Key Technologies R\&D Program of Jiangsu (Prospective and Key Technologies for Industry), No. BE2022067 and BE2022067-3, Natural Science Foundation of Jiangsu Province, No. BK20211182, open research fund of National Mobile Communications Research Laboratory, No. 2022D04, Key Project of Aeronautical Science Foundation of China, No. 2020Z073009001. ({\it Corresponding author: Qiuming Zhu, Cheng-Xiang Wang})}
\thanks{B. Hua, H. Ni, Q. Zhu, T. Zhou, K. Mao, and X. Zhang are with the Key Laboratory of Dynamic Cognitive System of Electromagnetic Spectrum Space, College of Electronic and Information Engineering, Nanjing University of Aeronautics and Astronautics, Nanjing 211106, China (e-mail: {byhua; nihaoran; zhuqiuming; zhoutongtong; maokai; zhangxiaofei}@nuaa.edu.cn).}
\thanks{C.-X. Wang is with the National Mobile Communications Research Laboratory, School of Information Science and Engineering, Southeast University, Nanjing 210096, China, and also with the Pervasive Communication Research Center, Purple Mountain Laboratories, Nanjing 211111, China (e-mail: chxwang@seu.edu.cn).}
\thanks{J. Bao is with the Department of Physics, Nanjing University of Aeronautics and Astronautics, Nanjing, 210016, China (e-mail: broadenway@nuaa.edu.cn).}
}

\maketitle
\vspace{-18mm}
\begin{abstract}
Unmanned aerial vehicle (UAV)-to-ground (U2G) channel models play a pivotal role in reliable communications between UAV and ground terminal. This paper proposes a three-dimensional (3D) non-stationary hybrid model including large-scale and small-scale fading for U2G multiple-input-multiple-output (MIMO) channels. Distinctive channel characteristics under U2G scenarios, i.e., 3D trajectory and posture of UAV, fuselage scattering effect (FSE), and posture variation fading (PVF) are incorporated into the proposed model. The channel parameters, i.e., path loss (PL), shadow fading (SF), path delay, and path angle, are generated incorporating machine learning (ML) and ray tracing (RT) techniques to capture the structure-related characteristics. In order to guarantee the physical continuity of channel parameters such as Doppler phase and path power, the time evolution methods of inter- and intra- stationary intervals are proposed. Key statistical properties, including temporal auto-correction function (ACF), power delay profile (PDP), level crossing rate (LCR), average fading duration (AFD), and stationary interval (SI), are analyzed with the impact of the change of fuselage and posture variation. It is demonstrated that both posture variation and fuselage scattering have crucial effects on channel characteristics. The validity and practicability of the proposed model are verified by comparing the simulation results with the measured ones.
\end{abstract}

\section{Introduction}
\IEEEPARstart{W}{ith} emerging global connectivity requirements, the space-air-ground-sea integrated networks are proposed to provide seamless coverage, ultra-reliable connection and user diversity\cite{WangCX20VTM}. Owing to the low cost, high mobility and versatility, unmanned aerial vehicles (UAVs) have been considered to be a promising paradigm for establishing air-ground communication networks\cite{ChengX19JCIN}. Thus, the research of UAV-to-ground (U2G) wireless communication has become an essential part of the pre-research work of next generation mobile communication systems\cite{Khawaja19CST}. Unlike terrestrial communication nodes, UAVs have unique characteristics, including three-dimensional (3D) arbitrary trajectory, 3D antenna arrangement, and 3D rotational posture \cite{AiB20VTC}. These new features would cause different non-stationarity that conventional channel models cannot precisely present. To better design and evaluate future air-ground communication systems, it is critical to provide a realistic and reliable U2G channel model\cite{Ullah20TCCN}.
\vspace{-1mm}
\subsection{Related Works}
From the perspective of modeling approaches, modern U2G channels can be modeled by the statistical method, geometrical method, and machine learning (ML)-enabled method \cite{HeRS22TAP}. The statistical channel models employ mathematical formulas from empirical observations such as field measurements or ray-tracing (RT) simulations. These models are usually employed for link-level analysis due to the lack of implicit temporal and spatial continuity. The shortcomings of pure statistical models are solved by incorporating geometrical information, and the improved models are usually referred to as geometry-based stochastic channel models (GBSMs). In fact, many standardized channel models utilize the geometrical method assuming different geometry shapes or unique channel characteristics \cite{3GPP20}. By assuming that the scatterers obey diverse 3D geometric distributions, such as cylinder \cite{AiB20TVT}, ellipsoid \cite{ZhangZC18CL}, and sphere \cite{ChengX17VTC}, GBSMs can support 3D propagation scenarios and become the mainstream method of U2G channel modeling. The fundamental concept of the ML-enabled method is obtaining the latent channel characteristics using ML technologies \cite{Oroza17TCCN, ZhaoXW19CNCOM, WangCX19CM}. However, the ML method can merely provide channel information at a particular condition and heavily rely on the training dataset. Therefore, the future U2G channel modeling may combine various methods for their advantages.

Several U2G channel models were proposed to describe the motion characteristic of UAVs. For example, a basic 3D GBSM was proposed in \cite{Gulfam16AS} to study the U2G channels, which assumed the ground terminals were fixed. However, that assumption limited the versatility of such kind of models. Some existing U2G channel models assumed that the UAV or ground terminal moved at constant speed in a straight line \cite{GuanK19VTC}. But in a practical environment, the vehicle may experience velocity change and arbitrary trajectory. The U2G channel models in \cite{Borhani17TVT} considered both the UAV and vehicle moved with constant velocities, which was inconsistent with common scenes. A modified U2G channel model considering the 3D speed variation of UAV was proposed in \cite{ChengX19IETCom}, but the trajectory of the ground terminal was still fixed. By setting the trajectory of both UAV and ground terminal, authors in \cite{GuanK19AWPL} proposed a more realistic dual-mobile UAV channel model and studied the corresponding simulations. Moreover, the authors in \cite{ZhuQM19MAP} proposed a U2G multiple-input multiple-output (MIMO) channel model, and the transceivers moved along 3D trajectories.

On the other hand, U2G channels have obvious non-stationary characteristics when UAVs move rapidly in 3D space. The researchers upgraded the stationary channel models by introducing time-variant channel parameters. For example, authors in \cite{ChengX19IoTJ} introduced time-varying departure and arrival angle parameters into the concentric cylindrical channel model, and the modified model can reflect the non-stationary characteristics of the UAV channel. A non-stationary MIMO U2G channel model was proposed in \cite{AiB21JSAC}, where the distribution of time-variant scatterers was improved to support the non-stationarity. To describe the non-stationary effect, authors in \cite{WangCX19Access} employed the geometric relationship between the sphere and truncated ellipsoid in the U2G channel model to update the time-variant channel parameters dynamically.
\vspace{-3mm}
\subsection{Motivations}
The aforementioned U2G channel models are not realistic enough because they merely focus on the motion characteristic of UAVs. However, some other characteristics of UAV, such as posture rotation and fuselage scattering, are also pivotal \cite{JangICCAS15}. The drone pitch rotation and its impact on the channel were studied in \cite{AiB20TVT2}, thus, the model can describe the basic posture rotation. Further, the effects of three posture angles and fuselage scattering were discussed in \cite{ZhuQM21CSPS}, and a modified U2G channel model was proposed. Besides, a U2G channel model considering the visible region of the UAV antenna was proposed in \cite{JiangLG21IoTJ}, and the birth-death procedure of the local multi-path components was analyzed. The authors in \cite{WangCX21TVT} combined the channel model with the airborne random motion model, where the UAV trajectory related to time-variant heading was modified. A U2G channel model with dynamic space-time cluster parameters was proposed in \cite{ChengX22TWC}, and the space-time clusters were analyzed in the antenna array domain and time domain. The RT method was utilized to achieve a more realistic channel description in \cite{ZhuQM20Sensors}, and a U2G channel model considering antenna beamforming was established. These concepts, which incorporate more deterministic information into U2G channel modeling, will also be applied in this paper to ensure the authenticity of U2G channel modeling.

Meanwhile, it should be noticed that some channel parameters, e.g., the Doppler frequency shift and path power, in non-stationary models are not realistic enough. For example, authors in \cite{WangCX17TWC, AiB18TWC} defined the time-variant Doppler frequency shift instead of the constant one. However, it leads to physically inconsistent phase variations in the channel model \cite{Patzold17TWC}. The concept of calculating Doppler frequency by the integral method was mentioned in \cite{ZhuQM18TC} to get a more realistic phase variation. The path power parameter was discussed in \cite{WangCX21IoTJ}, where the time-variant delays were employed to calculate the power for each sub-path. Moreover, authors in \cite{QuaDRiGa21} pointed out that the path phase should be determined by total path length to capture the Doppler effect caused by dual-mobility, and a squared sine function could model the path power ramp. However, a thorough channel parameter generation procedure is still absent for establishing a more realistic U2G non-stationary channel model.

The limitations mentioned above suggest a requirement to develop a non-stationary U2G channel model considering the unique characteristics of UAVs with more realistic channel parameter generation and time-evolution. This paper aims to fill these research gaps.
\vspace{-3mm}
\subsection{Contributions}
Motivated by the above background and current research gaps, this paper proposes a realistic hybrid channel model combining the geometrical method, RT approach, and ML technology to consider the deterministic information and unique characteristics of the U2G communication scenario. The contributions and novelties of this article include the following.

1) \textcolor{red}{A non-stationary wideband U2G MIMO channel model is presented by thoroughly considering the impact of UAV posture variation and fuselage scattering effect (FSE).} The model describes both large-scale and small-scale fading in three segments, i.e., near-UAV segment (NUS), free space link (FSL), and near-ground segment (NGS). The posture variation fading (PVF) coefficient and the deterministic scattering caused by FSE are considered. Based on these improvements, the realistic U2G channel can be characterized by the proposed model.

2) Generation and time-evolution procedure of segmented time-variant channel parameters, including the path loss (PL), shadow fading (SF), delay, power, angle, and phase, are given. The RT method and ML technology are employed to modify the traditional empirical and geometry-based parameter calculation method. By introducing the jerk-limited function to describe the path power ramp and utilizing the propagation distance variation to capture the Doppler effect, the time-evolution continuity of the channel parameters is achieved, which describes the channel non-stationarity more accurately.

3) Statistical properties of the proposed U2G channel model are given and simulated. The simulated channel characteristics, including temporal auto-correction function (ACF), power delay profile (PDP), level crossing rate (LCR), average fading duration (AFD), and stationary interval (SI) are analyzed and compared with results from other channel models, which verifies the effectiveness of the proposed model.

4) The practicability of the proposed model is validated by measurement data. Measured results from the measurement campaigns or referred literature are compared with simulation results of the proposed model, including PL, ACF, SI, and PVF coefficient. Moreover, the PL, ACF and PVF coefficient show that UAV posture variation and fuselage scattering have vital impacts on channel characteristics.

The remainder of this paper is organized as follows. Section~II describes a 3D non-stationary U2G channel model incorporating posture variation and FSE. Section~III studies the generation and time-evolution of the model parameters. Statistical properties of proposed model are presented and analyzed in Section~IV. Section~V compares and discusses the analytical, simulated, and measured results. Conclusions are finally drawn in Section~VI.
\section{Channel Model Incorporating Posture Variation and FSE}
\begin{figure}
\centering
\includegraphics[height=8cm]{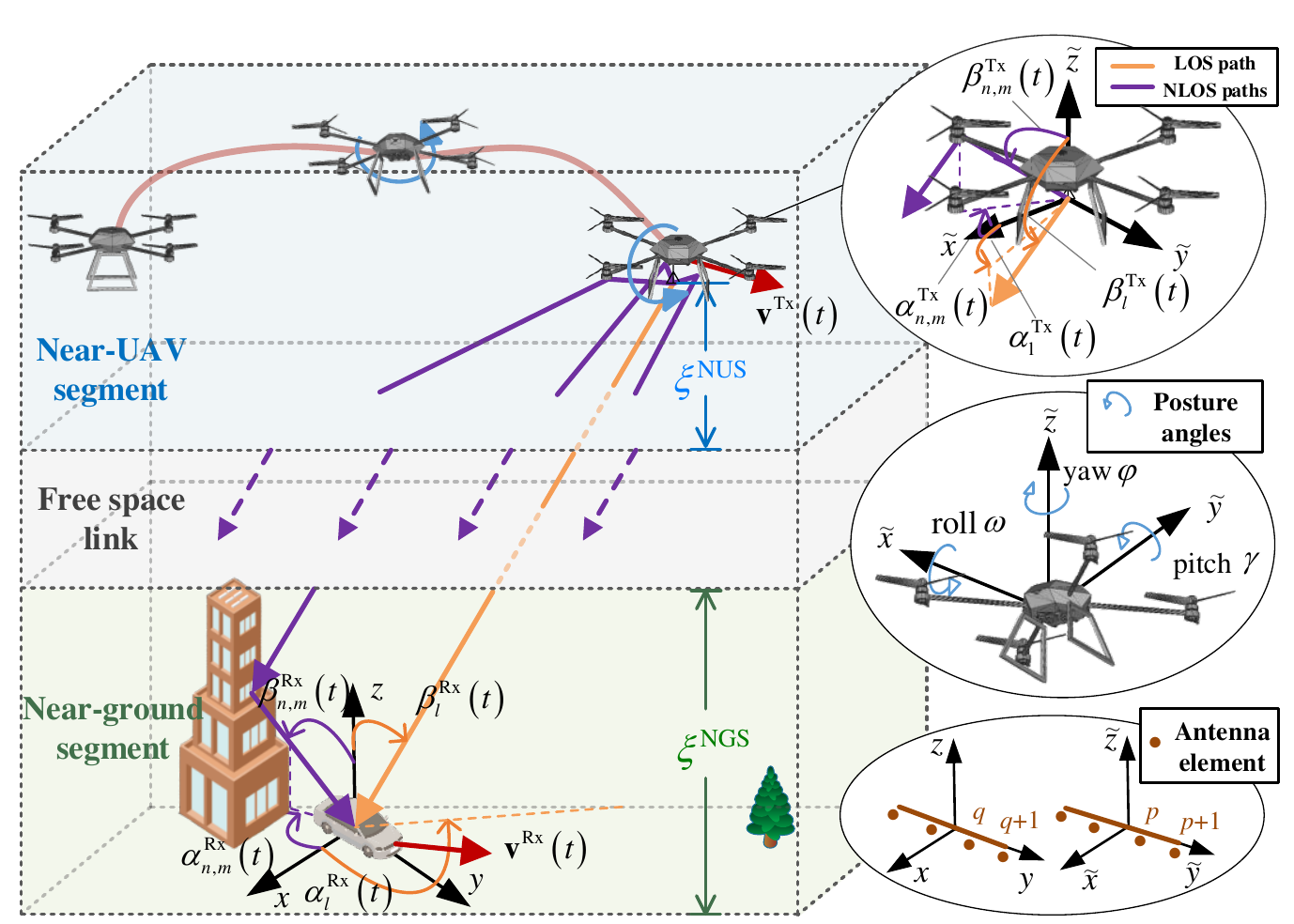}
\vspace{-2mm}
\caption{\color{red}Typical U2G communication scenario.}\label{scenario}
\vspace{-3mm}
\end{figure}

A typical U2G communication scenario is shown in Fig.~\ref{scenario}. The transmitter (Tx) on the UAV equips with $P$ antenna elements, and the receiver (Rx) on the ground vehicle equips with $Q$ antenna elements. Note that $\widetilde{x}$\rm{-}$\widetilde{y}$\rm{-}$\widetilde{z}$ and $x$\rm{-}$y$\rm{-}$z$ are two independent coordinate systems for Tx and Rx, respectively. The origin position of the coordinate system is located on the first element of the equipped antenna array. The communication channel consists of the line-of-sight (LoS) component and non-LoS (NLoS) components. Moreover, posture angles for pitch, roll, and yaw are defined to describe the time-variant posture variation of UAV.
\textcolor{red}{
For clarity, the time-variant roll, yaw, and pitch angles are marked as $\omega$, $\varphi$, and $\gamma$, respectively. Roll angle rotates around $\widetilde{x}$ axis and $\omega \in \left( -\pi,\pi  \right]$, pitch angle rotates around $\widetilde{y}$ axis with respect to the horizontal plane and $\gamma \in \left( -\pi,\pi  \right]$, and yaw angle rotates around $\widetilde{z}$ axis in the horizontal plane with a counter-clockwise increasing mode and $\varphi \in \left[ 0,2\pi  \right)$. Furthermore, the origin position of the coordinate system is located on the first element of the equipped antenna array.}
The detailed parameters are listed in Table~\ref{table1}.

\begin{table}
\caption{DEFINITION OF KEY CHANNEL MODEL PARAMETERS\label{table1}}
\centering
\setlength{\tabcolsep}{3pt}
\begin{tabular}{|m{100pt}<{\centering}|m{340pt}<{\centering}|}
\hline
Symbol & Definition\\
\hline
\small
${{\boldsymbol{v}}^i}(t)$, 
\scriptsize
$i \in \{ \rm{Tx},\rm{Rx} \}$ & 3D velocity vectors of the Tx and Rx\\
\hline
\footnotesize
${{\bf{L}}^i}(t)$, ${\bf{L}}_{n,m}(t)$, 
\scriptsize
$i \in \{ {\rm{Tx}},\rm{Rx} \}$ & Location vectors of the Tx, Rx, and NLoS scatterers\\
\hline
\small
$\alpha _l^i(t)$, $\beta _l^i(t)$, 
\scriptsize
$i \in \{ \rm{Tx},\rm{Rx} \}$ & Azimuth and elevation angle of departure (AAoD and EAoD) or arrival (AAoA and EAoA) in the LoS path\\
\hline
\small
$\alpha _{n,m}^i(t)$, $\beta _{n,m}^i(t)$, 
\scriptsize
$i \in \{ \rm{Tx},\rm{Rx} \}$ & AAoD and EAoD, or AAoA and EAoA in NLoS paths\\
\hline
\small
$\varphi (t)$, $\omega (t)$, $\gamma (t)$ & Time-variant posture agnles for pitch, roll, and yaw\\
\hline
\small
$\theta _{\rm{H},\vartheta}$,
\scriptsize
$\vartheta  \in \{ \varphi,\omega,\gamma \}$ & Half-power beam width of the UAV antenna projected on each axis\\
\hline
\small
$\xi ^{\rm{NUS}}$, $\xi ^{\rm{NGS}}$, $h^{{\rm{Tx}}}$ & Vertical distances for NUS and NGS, and height of the Tx on UAV\\
\hline
\small
$\tau _{pq}^{{\rm{LoS}}}(t)$, $\tau_{pq,n,m}^{\rm NLoS}(t)$ & Delay of the LoS path and NLoS paths\\
\hline
\footnotesize
$K(t)$, $C^{\rm{P}}(t)$, $P_{pq,n,m}^{\rm{NLoS}}(t)$ & Ricean factor, PVF coefficient, and normalized NLoS power\\
\hline
\small
$\Phi_{pq}^{\rm LoS}\left( t \right)$, $\Phi_{pq,n,m}^{\rm NLoS}\left( t \right)$ & Phase of the LoS path and NLoS paths\\
\hline
\end{tabular}
\vspace{-2mm}
\end{table}
\textcolor{red}{
In most U2G channel models, an aircraft is usually regarded as a point without the shape and posture for simplicity. However, ignoring FSE and time-varying posture may lead to distortion of channel characteristics. In addition, the channel characteristics in the free space segment can be considered separately. According to the different scattering environments, the propagation scenarios is divided into three parts, i.e., NUS, FSL, and NGS.}

The entire communication link includes one LoS path and numerous NLoS paths. The former represents a direct connection from Tx to Rx without any obstacles, while the latter experienced three different transmission links. Due to the FSE in NUS, part of the transmitted signal is first reflected by the fuselage and then reaches FSL. Then, the propagation direction of the NLoS path in FSL is assumed to be the same as that of the LoS path until the signal reaches the NGS. Finally, the NLoS path signal in NGS is reflected to Rx through its surrounding clusters.

The U2G MIMO fading channel between UAV and the ground terminal is affected by both large-scale fading and small-scale fading, and it can be defined as
\begin{equation}
\small
{{\bf{H}}_{PQ}}={{\left[ PL\cdot SF \right]}^{1/2}}\cdot{{\left[{{h}_{pq}}(t,\tau )\right]}_{P\times Q}}
\end{equation}
\textcolor{red}{where variable $PL$ represents the PL with respect to the propagation distance, angle, observing height, and fuselage structure}, and variable $SF$ denotes SF due to obstacles in the path of propagation. PL and SF are usually referred to as large-scale fading parameters and will be discussed in detail in the next chapter. The channel impulse response (CIR) between the $p$-th transmitting antenna element and the $q$-th receiving antenna element is represented as ${{h}_{pq}}(t,\tau )$, which can be expressed as the superposition of the LoS and NLoS components, i.e.,
\begin{equation}
\small
\textcolor{red}{
{h_{pq}}(t,\tau)\!=\! \sqrt{\frac{{K(t)}}{{K(t)\!+\!1}}} h_{pq}^{{\rm{LoS}}}(t) \delta \left( {\tau \!-\! \tau _{pq}^{{\rm{LoS}}}(t)} \right) \!+\! \sqrt {\frac{1}{{K(t)\!+\!1}}}\sum\limits_{n = 1}^{N(t)} \sum\limits_{m = 1}^M {h_{pq,n,m}^{{\rm{NLoS}}}(t) \delta \left({\tau \!-\! \tau _{pq,n,m}^{{\rm{NLoS}}}(t)}\right)} }
\label{eq.CIR}
\end{equation}
\textcolor{red}{
where $K(t)$ denotes the time-variant Rice factor, $N(t)$ are the total number of effective scattering paths, $M$ is the number of sub-paths within the $n$-th NLoS path, $\tau_{pq}^{\rm{LoS}}(t)$ and $\tau_{pq,n,m}^{\rm{NLoS}}(t)$ denote the delays for the LoS and NLoS path, respectively, the channel coefficient $h_{pq}^{\rm{LoS}}(t)$ and $h_{pq,n,m}^{\rm{NLoS}}(t)$ in (2) can be further modeled as}
\begin{equation}
\small
\textcolor{red}{
h_{pq}^{{\rm{LoS}}}(t) = {C^{\rm{P}}}(t) 
{\left[\!\!\!{\begin{array}{*{20}{c}}
{F_{p{\rm{,V}}}^{{\rm{Tx}}}\left( {{\boldsymbol{s}}_l^{{\rm{Tx}}}(t),{{\bf{R}}^{\rm{P}}}(t)} \right)}\\
{F_{p{\rm{,H}}}^{{\rm{Tx}}}\left( {{\boldsymbol{s}}_l^{{\rm{Tx}}}(t),{{\bf{R}}^{\rm{P}}}(t)} \right)}
\end{array}}\!\!\!\right]^{\rm{T}}}\!\!\left[\!\!\!{\begin{array}{*{20}{c}}
{F_{p{\rm{,V}}}^{{\rm{Rx}}}\left( {{\boldsymbol{s}}_l^{{\rm{Rx}}}(t),{{\bf{R}}^{\rm{P}}}(t)} \right)}\\
{F_{p{\rm{,H}}}^{{\rm{Rx}}}\left( {{\boldsymbol{s}}_l^{{\rm{Rx}}}(t),{{\bf{R}}^{\rm{P}}}(t)} \right)}
\end{array}}\!\!\!\right] \!\!{e^{{\rm{j}}\Phi _{pq}^{{\rm{LoS}}}(t)}} }
\label{eq.h_LoS}
\end{equation}
\begin{equation}
\small
\textcolor{red}{
h_{pq,n,m}^{{\rm{NLoS}}}(t) \!=\! {C^{\rm{P}}}(t) 
{\left[\!\!\!{\begin{array}{c}
{F_{p{\rm{,V}}}^{{\rm{Tx}}}\left( {{\boldsymbol{s}}_{n,m}^{{\rm{Tx}}}(t),{{\bf{R}}^{\rm{P}}}(t)} \right)}\\
{F_{p{\rm{,H}}}^{{\rm{Tx}}}\left( {{\boldsymbol{s}}_{n,m}^{{\rm{Tx}}}(t),{{\bf{R}}^{\rm{P}}}(t)} \right)}
\end{array}}\!\!\!\right]^{\rm{T}}}\!\!\left[\!\!\!{\begin{array}{c}
{F_{p{\rm{,V}}}^{{\rm{Rx}}}\left( {{\boldsymbol{s}}_{n,m}^{{\rm{Rx}}}(t),{{\bf{R}}^{\rm{P}}}(t)} \right)}\\
{F_{p{\rm{,H}}}^{{\rm{Rx}}}\left( {{\boldsymbol{s}}_{n,m}^{{\rm{Rx}}}(t),{{\bf{R}}^{\rm{P}}}(t)} \right)}
\end{array}}\!\!\!\right]
\!\!\!{\sqrt {P_{pq,n,m}^{{\rm{NLoS}}}(t)} {e^{{\rm{j}}\Phi _{pq,n,m}^{{\rm{NLoS}}}(t)}}}  }
\label{eq.h_NLoS}
\end{equation}
\textcolor{red}{
where $C^{\rm{P}}(t)$ is the posture-related power coefficient for describing the PVF, $F_{p/q{\rm{,V}}}^{{\rm{Tx/Rx}}}( \cdot )$ and $F_{p/q{\rm{,H}}}^{{\rm{Tx/Rx}}}( \cdot )$ denote the antenna pattern components in the vertical and horizontal planes, respectively, ${\boldsymbol{s}}_l^{\rm{Tx/Rx}}(t)$ and ${\boldsymbol{s}}_{n,m}^{{\rm{Tx/Rx}}}(t)$ are the angle unit vectors of the LoS and NLoS paths at the Tx (or Rx), which can be obtained by converting angles of departure (AoDs) and angles of arrival (AoAs) to Cartesian coordinates \cite{ZhuQM19MAP}. In (\ref{eq.h_NLoS}), $P_{pq,n,m}^{\rm{NLoS}}(t)$ is the normalized relative power coefficient of the $m$-th sub-path within $n$-th path, $\Phi_{pq}^{\rm{LoS}}(t)$ and $\Phi_{pq,n,m}^{\rm{NLoS}}(t)$ are the phase shifts of LoS and NLoS paths, respectively. In order to describe the posture effect, we introduce a posture matrix ${\bf{R}}^{\rm{P}}(t)$ to keep the coordinate system consistent before and after the UAV rotation, which can be expressed as}
\begin{equation}
\small
\textcolor{red}{
{{\bf{R}}^{\rm{P}}}(t) = \left[ {\begin{array}{*{20}{c}}
{\cos \varphi(t) }&{ - \sin \varphi(t) }&0\\
{\sin \varphi(t) }&{\cos \varphi(t) }&0\\
0&0&1
\end{array}} \right]\left[ {\begin{array}{*{20}{c}}
{\cos \gamma(t) }&0&{\sin \gamma(t) }\\
0&1&0\\
{ - \sin \gamma(t) }&0&{\cos \gamma(t) }
\end{array}} \right]\left[ {\begin{array}{*{20}{c}}
1&0&0\\
0&{\cos \omega(t) }&{ - \sin \omega(t) }\\
0&{\sin \omega(t) }&{\cos \omega(t) }
\end{array}} \right] . }
\label{eq.r^p}
\end{equation}
\textcolor{red}{
Based on these time-varying parameters, the proposed model can more realistically describe the propagation characteristics of U2G channels.
}

\section{Channel Parameters Generation and Time-Evolution}
\begin{figure}
\centering
\includegraphics[height=8cm]{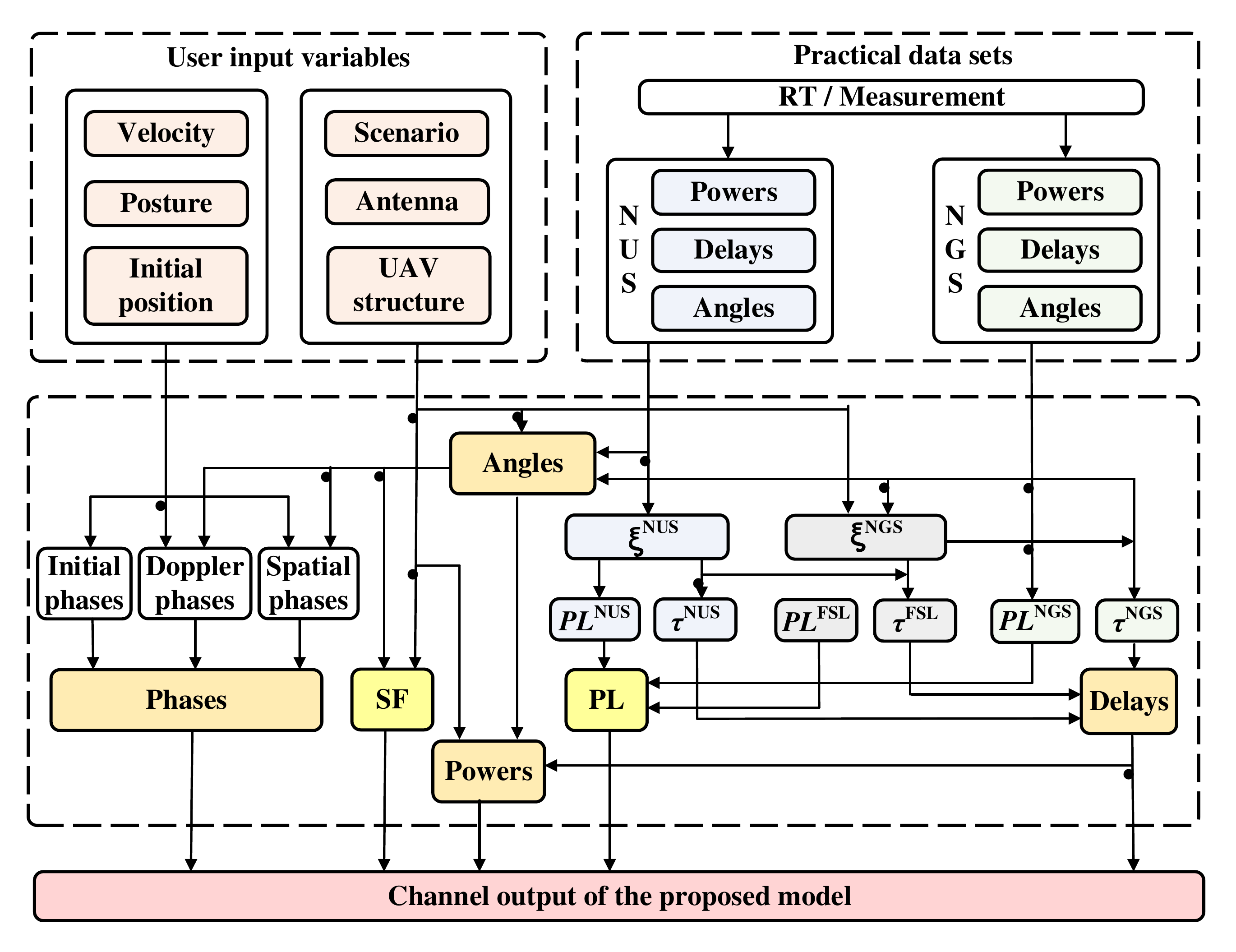}
\caption{Channel generation flow chart of the proposed model.}\label{Channel generation}
\vspace{-5mm}
\end{figure}

The generation and time-evolution of channel parameters are crucial to the channel model. By comprehensively considering 3D scattering space, 3D arbitrary trajectory, 3D antenna layout, and 3D rotational posture, the corresponding channel parameters can be generated more realistically, including PL, path angle, path phase, path delay, and path power. Fig.~\ref{Channel generation} shows the channel generation flow chart of the proposed model.

\vspace{-1mm}
\subsection{Path Loss and Shadow Fading}
\textcolor{red}{
Due to the different propagation environment in the NUS, FSL, and NGS, PL not only varies with distance, but also is highly correlated with height. For example, PL in NGS is affected by scatterers such as buildings and trees, while PL in NUS is only affected by the fuselage structure. The traditional PL model cannot capture this feature. This paper uses deterministic data, empirical formula and ML methods to obtain the ${{PL}^{\rm{NUS}}}$ in the NUS, ${{PL}^{\rm{FSL}}}$ in the FSL, and ${{PL}^{\rm{NGS}}}$ in the NGS, respectively.}

\textcolor{red}{
In the NUS, ${{PL}^{\rm{NUS}}}$ is determined by fuselage structure and propagation distance. Therefore, it can be obtained by deterministic approaches, such as field measurement or RT method. The RT method is chosen in this step because it can provide massive data to support model analysis more conveniently than field measurement. In addition, the RT procedure can also provide the propagation length in the NUS, which will be employed in calculation of delay and phase parameters.}

\textcolor{red}{
The observing height $h$ in the NUS satisfies condition ${h^{{\rm{Tx}}}} - {\xi ^{{\rm{NUS}}}} \le h$. The distance $\xi ^{{\rm{NUS}}}$ can be determined by the constraint that the fuselage scattering having a sufficient effect on PL. The PL considering the physical structure will have 10\% additional loss compared with the unconsidered case at $\xi ^{{\rm{NUS}}}$ below the UAV. Several types of UAVs, such as fixed-wing UAV, quadrotor UAV, hexacopter UAV, have been simulated, and the results show that $\xi ^{{\rm{NUS}}}$ values are ranging from a few tens to a hundred meters.}

By reproducing all rays between Tx and Rx, specific rays are selected according to the electric field strength to represent the NLoS paths of signal propagation. For the $m$-th reflection ray, i.e. the $m$-th sub-path in proposed model,
\textcolor{red}{the electric field vector can be calculated by\cite{Rappaport02}}
\begin{equation}
\small
{\bf{E}}_{n,m}(d_{n,m})=\frac{ {{\bf{E}}_1} R {{e}^{\rm{-j}2\pi\it{d_{n,m}}/{\lambda}} } } {d_{n,m}}
\label{eq.e_nm_RT}
\end{equation}
\textcolor{red}{
where $d_{n,m}$ is the propagation distance of $m$-th ray, ${\bf{E}}_1$ is the unit electric vector, $R$ is reflection coefficient, $\lambda$ is signal wavelength. All rays are superimposed coherently at a specified distance, and the total electric field intensity of the receiving point can be expressed as\cite{Rappaport02} }
\begin{equation}
\small
E^{\rm{RT}}(d)= \left| \sum\limits^{N(d)}_{n=1} \sum\limits^{M(d)}_{m=1}{{\bf{E}}_{n,m}(d_{n,m})}\right|
\end{equation}
where $N(d)$ and $M(d)$ denotes the number of paths and sub-paths at the point where propagation distance is $d$ in RT procedure, respectively. Finally, PL is the ratio of transmit power to received power, thus $PL^{\rm{NUS}}$ can be expressed as
\begin{equation}
\small
PL^{\rm{NUS}}(d) = 20\log_{10} \left( {E_0}/{E^{\rm{RT}}(d)} \right)
\end{equation}
where $E_0$ is the total electric field intensity of emitting rays.

There is no scattering in the FSL, thus ${PL}^{\rm{FSL}}$ can be acquired by the empirical formula
\begin{equation}
\small
PL^{\rm{FSL}}(d)=20\log_{10} d+20\log_{10}{f_{c}}+32.4
\label{eq.pl^fsl}
\end{equation}
where ${{f}_{c}}$ is frequency of the carrier wave.  
\textcolor{red}{Note that the constant is 32.4 when the unit of frequency is GHz and the unit of distance is m.}

In the NGS, the characteristics of ${PL}^{\rm{NGS}}$ are close to those of traditional ground scenarios, but the Tx is assumed at the height of $\xi^{\rm{NGS}}$.
\textcolor{red}{
The parameter $\xi^{\rm{NGS}}$ is highly-related to the height of ground buildings, and can be set to different values depending on scenarios. Typical values of building height are listed in the ITU-R urban deployment model \cite{ITU-R09}. For suburbs, cities, dense cities, and high-rise cities, the average building height is 8~m, 15~m, 20~m, and 50~m, respectively.}

In order to obtain a more accurate PL in NGS, this paper introduces the ML algorithm into the PL model. Firstly, by fixing Tx at the height $\xi^{\rm{NGS}}$, a large number of propagation distance $d$ and AoA data are obtained by RT method. 
\textcolor{red}{Then, a back propagation neural network (BPNN) based on ${PL}^{\rm{NGS}}$, $d$, $\alpha$ and $\beta$ is constructed. Note that the input training data contains the information about the scattering environment near the receiver. By using the internal relationship between the data and the parameters such as distance and AoA, PL can be predicted as}
\begin{equation}
\small
PL^{\rm NGS}\left( {d,\alpha,\beta } \right) = {\rm{F}_{\rm{BPNN}}} \left( {d,\alpha,\beta; {\boldsymbol{w}},{\boldsymbol{b}},{\boldsymbol{\sigma}}} \right)
\end{equation}
where $\alpha$ and $\beta$ denote the azimuth and elevation angle, ${{\rm{F}}_{\rm{BPNN}}}(\cdot )$ is the transfer function of BPNN determined by the network, $\boldsymbol{w}$, $\boldsymbol{b}$ and $\boldsymbol{\sigma}$ are the weight matrix, bias matrix, and activation matrix of BPNN, respectively.
\textcolor{red}{
After obtaining the original data set {\small $\left\{ {PL_{n,m}^{{\rm{NGS}}},{d_{n,m}},\alpha _{n,m}^{{\rm{Rx}}},\beta _{n,m}^{{\rm{Rx}}}} \right\}$}, the elements in the data set are divided into training set and validation set according to the ratio of 7:3. The training set {\small $\left\{ {PL_{n,m}^{{\rm{NGS,tr}}},d_{n,m}^{{\rm{tr}}},\alpha _{n,m}^{{\rm{Rx,tr}}},\beta _{n,m}^{{\rm{Rx,tr}}}} \right\}$} is used to train the parameters of neural network, and the validation set {\small $\left\{ {PL_{n,m}^{{\rm{NGS,va}}},d_{n,m}^{{\rm{va}}},\alpha _{n,m}^{{\rm{Rx,va}}},\beta _{n,m}^{{\rm{Rx,va}}}} \right\}$} is used to verify whether the trained network has over-fitting or under-fitting problems.}
Taking a neural network with an input layer, a hidden layer with $\chi$ neurons, and an output layer as an example, the PL output prediction model is expressed as
\begin{equation}
\small
\begin{aligned}
& {\rm{F}_{\rm{BPNN}}} \left( {d,\alpha,\beta; {\boldsymbol{w}},{\boldsymbol{b}},{\boldsymbol{\sigma}}} \right) =\sigma _{1}^{(2)}[\sum\limits_{i=1}^{\chi}{w_{i1}^{(2)}\sigma_{j}^{(1)}(w_{1i}^{(1)}d_{n,m}^{\rm{tr}}+b_{i}^{(1)})}+b_{1}^{(2)}]\\
& +\sigma _{2}^{(2)}[\sum\limits_{i=1}^{\chi}{w_{i1}^{(2)}\sigma _{j}^{(1)}(w_{1i}^{(1)}\alpha_{n,m}^{\rm{Rx,tr}}+b_{i}^{(1)})}+b_{2}^{(2)}]+\sigma_{3}^{(2)}[\sum\limits_{i=1}^{\chi}{w_{i1}^{(2)}\sigma_{j}^{(1)}(w_{1i}^{(1)}\beta_{n,m}^{\rm{Rx,tr}}+b_{i}^{(1)})}+b_{3}^{(2)}]
\end{aligned}
\end{equation}
where $w_{ij}^{(k)}\in \boldsymbol{w}$ is the connection weight between the $j$-th neuron in the $k$-th layer and the $i$-th neuron in the former layer, $b_{j}^{(k)}\in \boldsymbol{b}$ is the bias of the $j$-th neuron in the $k$-th layer, and $\sigma_{j}^{(k)}(\cdot)\in \boldsymbol{\sigma}$ is the activation function of the $j$-th neuron in the $k$-th layer. In this paper, ${{\sigma }_{\rm hid}}\left( x \right)={1}/(1+{{e}^{-x}})$ is applied in the hidden layer, and ${{\sigma}_{\rm out}}\left( x \right)=\max \left( x,0 \right)+{{s}_{n}} \min \left( 0,x \right)$ is applied in the output layer, where ${{s}_{\rm n}}$ is a nonzero slope.

\textcolor{red}{
Based on the above analysis, the segmented PL can be modeled by}
\begin{equation}
\color{red}{
\small
PL\left( {d,h,\alpha,\beta } \right) = \left\{ {\begin{array}{*{20}{c}}
{20\log_{10} \left( {E_0}/{E^{\rm{RT}}(d)} \right)}&{{h^{{\rm{Tx}}}} - {\xi ^{{\rm{NUS}}}} \le h}\\
{20\log_{10} d + 20\log_{10} {f_c} + {C_1}}&{{\xi ^{{\rm{NGS}}}} \le h \le {h^{{\rm{Tx}}}} - {\xi ^{{\rm{NUS}}}}}\\
{{{\rm{F}}_{{\rm{BPNN}}}}\left( {d,\alpha,\beta ;{\boldsymbol{w}},{\boldsymbol{b}},{\boldsymbol{\sigma}}} \right) + {C_2}}&{h \le {\xi ^{{\rm{NGS}}}}}
\end{array}} \right.
}
\label{eq.PL}
\end{equation}
\textcolor{red}{
where $C_1$ and $C_2$ are constants to ensure the spatial consistency of PL calculations. When observing height satisfies $h = h^{\rm{UAV}} - \xi ^{\rm{NUS}}$ and $h = \xi ^{\rm{NGS}}$, the height-related break points can be obtained by $d_1 = {\xi^{\rm{NUS}}}/\sin \left( \beta_l^{\rm{Tx}}(t) \right)$ and $d_2 = {\xi^{\rm{NGS}}}/\sin \left( \beta_l^{{\rm{Rx}}}(t) \right)$, respectively. Note that the PL of LoS path can be regarded as a special case of (\ref{eq.PL}) and simplified as (\ref{eq.pl^fsl}). }

On the other hand, SF occurs when the signal is shadowed or blocked by obstacles between the Tx and Rx, which will lead to a reduction in signal strength. In the proposed model, SF only occurs in the NGS. SF can be modeled as a log-normal distribution according to the measurement results in various environments. The probability density function (PDF) of SF can be denoted as
\begin{equation}
\small
{{f}_{\rm{SF}}}(SF)=\frac{1}{\sqrt{2\rm{ }\pi\rm{ }}\sigma S{{F}_{{}}}}\exp \left[ -\frac{{{\left( \ln SF-{{\mu }_{{}}} \right)}^{2}}}{2\sigma _{{}}^{2}} \right], \ \ \ 0\le SF<\infty
\label{ShadowingPDF}
\end{equation}	
where $\sigma$ is the standard deviation of $SF$ in dB, and $\mu$ is the mean value of $SF$ in dB. 
\textcolor{red}{Note that the standard deviation and mean value of SF are determined by the scenario. Based on the channel measurements under the urban scenario, $\sigma$ is set as 8.1~dB, and $\mu$ is set as 19.5~dB\cite{Ho08EuCAP}.}
However, for different U2G scenarios, $\sigma$ and $\mu$ are not supposed to be a fixed value. In order to describe SF more accurately, this paper improves the parameter calculation method. RT method is employed to obtain a large amount $SF$ data under a specific scenario. Note that electromagnetic wave losses in RT simulations arise from interaction with the environment and atmospheric losses, thus we set atmosphere to vacuum during the simulation to obtain $SF$. On this basis, the ground environment and receiver position are fixed, and the UAV flies freely at the altitude of ${\xi}^{\rm NGS}$ in different positions without any posture variation. Then we preprocess the $SF$ data to fit the log-normal distribution, and obtain the $\sigma$ and $\mu$ corresponding to the given scenario. Finally, substitute the $\sigma$ and $\mu$ into (\ref{ShadowingPDF}), the PDF of $SF$ for the specific scenario can be obtained.
\vspace{-1mm}
\subsection{Path Delays and Powers}
Path delay describes the length of time that the signal reaches the Rx through different propagation paths. In the proposed model, the delay of the LoS component is assumed to be deterministic. In order to obtain the correct channel parameters for the MIMO system, the calculations must be done for each element of an array antenna. First, the instant location vector of Tx and Rx can be expressed as
\begin{equation}
\small
{{\bf{L}}^{{\rm{Tx/Rx}}}}(t) = {{\bf{L}}^{{\rm{Tx/Rx}}}}({t_0}) + \int_{{t_0}}^t {{{\boldsymbol{v}}^{{\rm{Tx/Rx}}}}\left( {t'} \right)dt'}
\end{equation}
where ${{\bf{L}}^{{\rm{Tx/Rx}}}}({t_0})$ denotes the initial location of Tx and Rx. Combined with topology information, the delay of LoS path can be expressed directly by the ratio of distance to wave speed as
\begin{equation}
\small
\color{red}{
\tau _{pq}^{\rm{LoS}}\left( t \right) = \| \!\left( {\bf{L}}^{\rm{Tx}}(t) + {\boldsymbol{e}}_{p}(t) \right) - \left( {\bf{L}}^{\rm{Rx}}(t) + {\boldsymbol{e}}_{q}(t) \right) \! \| /c   }
\end{equation}
\textcolor{red}{
where ${\boldsymbol{e}}_{p}(t)$ and ${\boldsymbol{e}}_{q}(t)$ are the vector pointing to the $p$-th and $q$-th antenna element from the centers of Tx and Rx, respectively, $c$ is the wave velocity\cite{QuaDRiGa21}.}

The delay of NLoS path ${{\tau }_{pq,n,m}}\left( t \right)$ is the sum of three parts, i.e., {\small $\tau _{pq,n,m}^{{\rm{NUS}}}(t)$}, {\small $\tau_{pq,n,m}^{{\rm{FSL}}}(t)$}, and {\small $\tau _{pq,n,m}^{{\rm{NGS}}}(t)$}. The delay in the NUS is determined by the fuselage and can be calculated by
\begin{equation}
\small
\tau_{pq,n,m}^{{\rm NUS}}\left( t \right) = d_{n,m}/c
\end{equation}
where $d_{n,m}$ can be found in (\ref{eq.e_nm_RT}) from the RT procedure. Due to all the signals are assumed to propagate along the direction of LoS path, thus the delays for all NLoS components in the FSL are the same, and can be determined by
\begin{equation}
\small
{{\tau}_{pq,n,m}^{\rm{FSL}}}(t)=\left( h^{\rm Tx}-{{\xi}_{\rm{NGS}}} \right) / \left( (\sin{\beta _{l}^{\rm{Tx}}(t)}) c \right)
\end{equation}
where the height of the Tx on UAV $h^{\rm Tx}$ is naturally greater than ${{\xi}_{\rm{NGS}}}$ in the U2G scenario. 

\begin{figure}
\centering
\includegraphics[height=6cm]{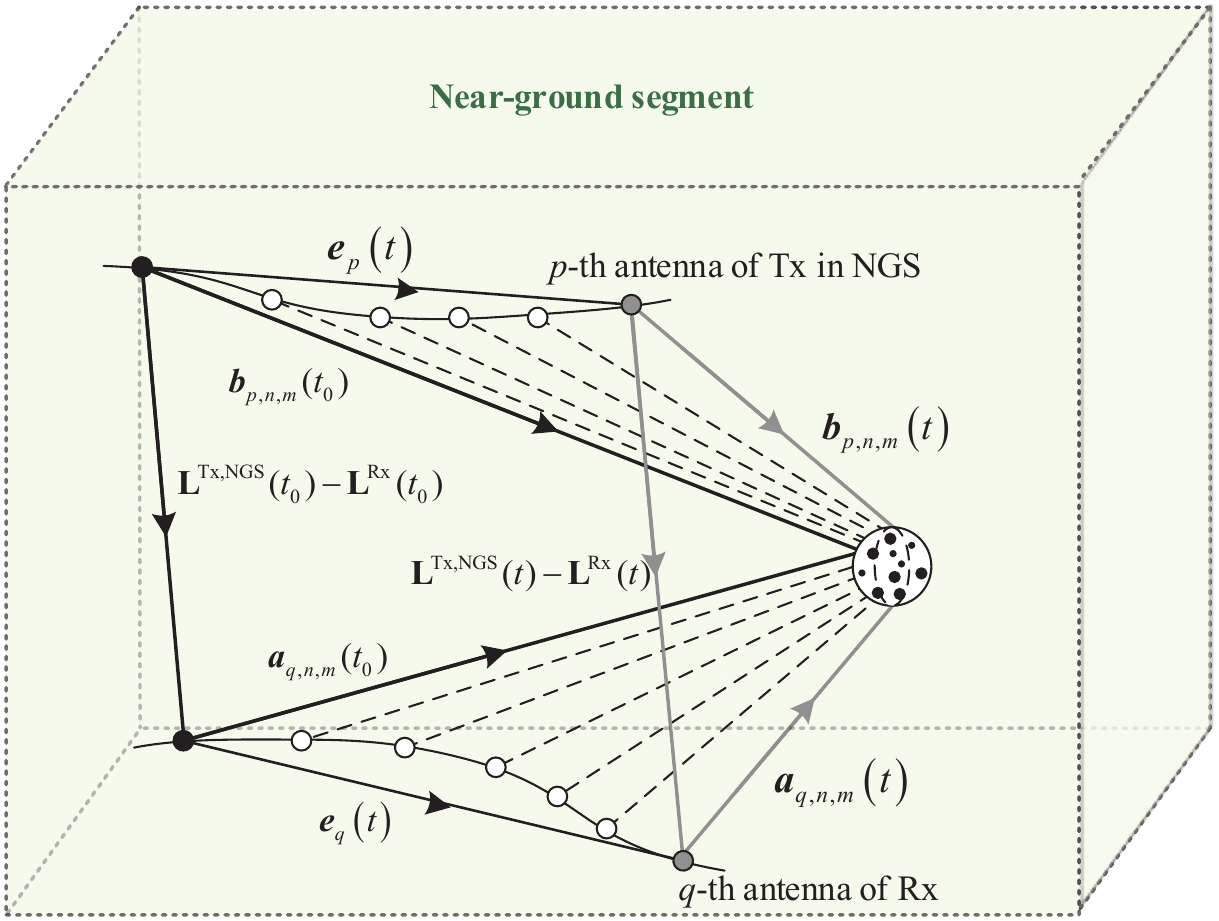}
\caption{Topology relationship during a stationary interval in the NGS.}\label{drifting}
\vspace{-5mm}
\end{figure}

The calculation of path delays in the NGS is based on the plane wave assumption where the location vector ${{\bf{L}}^{\rm{Tx,NGS}}}(t)$ is determined by the ${{\bf{L}}^{{\rm{Tx}}}}(t)$, ${{\bf{L}}^{{\rm{Rx}}}}(t)$, and ${{\xi}_{\rm{NGS}}}$. For different Tx positions with respect to the last-bounce scatterer (LBS), there is a triangle topology relationship during a stationary interval as shown in Fig.~\ref{drifting}.
The initial length of the NLoS path is generated from the measured delay spread, and the time-evolution process can be expressed as
\vspace{-1mm}
\begin{equation}
\small
\color{red}{
d _{pq,n,m}^{{\rm{NGS}}}\left( t \right) =  \left| {{\boldsymbol{b}}_{p,n,m}}(t) \right|+\left| {\boldsymbol{a}}_{q,n,m}(t) \right| }
\end{equation}
where ${\boldsymbol{b}}_{p,n,m}(t)$ and ${\boldsymbol{a}}_{q,n,m}(t)$ are the distance vector pointing to the LBS from the Tx and Rx, respectively. Moreover, at the moment $t$, the distance vectors are updated by geometric relations as follows
\vspace{-3mm}
\begin{equation}
\small
{{\boldsymbol{a}}_{q,n,m}}\left( t \right)={{\boldsymbol{a}}_{q,n,m}}\left( {{t}_{0}} \right)-{{\boldsymbol{e}}_{q}}\left( t \right)	
\end{equation}
\begin{equation}
\small
\color{red}{
{{\boldsymbol{b}}_{p,n,m}}\left( t \right)={{\bf{L}}^{\rm{Tx,NGS}}}(t_0)-{{\mathbf{L}}^{\rm{Rx}}}(t_0)+{{\boldsymbol{a}}_{q,n,m}}\left( t \right)-{{\boldsymbol{e}}_{p}}\left( t \right).}
\end{equation}
\textcolor{red}{
Based on these distance vectors, the length of the $m$-th sub-path within the $n$-th path in the NGS can be obtained, and the mean delay can be calculated by}
\begin{equation}
\small
\color{red}{
\tau _{pq,n,m}^{\rm{NGS}}\left( t \right) = \left( \left| {{\boldsymbol{b}}_{p,n,m}}\left( t \right) \right|+\left| {{\boldsymbol{a}}_{q,n,m}}\left( t \right) \right| \right)/c  .}
\end{equation}
\indent After all the delay parameters are generated, the power parameter of each path can be calculated. In the proposed model, the LoS path power is controlled by the time-variant Ricean factor $K(t)$, which describes the power ratio between LoS and NLoS paths. Then, the power allocation for each NLoS paths can be obtained by
\begin{equation}
\small
\tilde P_{pq,n,m}^{\rm NLoS}(t)=\exp \left(-{{\tau }_{pq,n,m}}\left( t \right)\frac{{{r}_{\tau }}-1}{{{r}_{\tau}}{{\sigma }_{\tau}}} \right){{10}^{-{{SF}_{\rm c}}/10}}
\end{equation}
where ${{r}_{\tau}}$ is the delay scalar, and ${{\sigma}_{\tau}}$ is the delay spread, ${{SF}_{\rm c}}$ is the cluster shadow fading in dB and obeys the Gaussian distribution \cite{WangCX18TCom}. \textcolor{red}{Finally, the powers of sub-paths need to be normalized, and the final relative power of NLoS path can be expressed as}
\begin{equation}
\small
\color{red}{
P_{pq,n,m}^{\rm{NLoS}}(t) = \tilde P_{pq,n,m}^{\rm{NLoS}}(t)/\sum\limits_{n=1}^{N(t)}\sum\limits_{m=1}^{M}{\tilde P_{pq,n,m}^{\rm{NLoS}}(t)}  .}
\end{equation}
\indent Based on the above expression, it can be found that the time-variant power changes with the delay parameter. Note that the location vectors and the geometric relations shown in Fig.~\ref{drifting} are valid only in the wide-sense stationary condition. The stationary interval determines the length of a stationary segment, which will be discussed in the next chapter. At the beginning of each stationary segment, the initial delay will be generated in a stochastic approach, which leads to the inconsistency of power variation. To deal with that, transitions of path power between the adjacent stationary segments are modeled by a squared sine function as 
\begin{equation}
\small
\color{red}{
w^{[\rm{sin}]} = \sin^{2} \left( \pi w^{[\rm{lin}]}/2 \right)  }
\end{equation}
where $w^{[\rm{lin}]}$ is the linear ramp ranging from 0 to 1, and $w^{[\rm{sin}]}$ is the corresponding sine-shaped ramp with a constant slope at the beginning and the end, more detail can be found in \cite{QuaDRiGa21}.

Additionally, the radiation direction of antenna may point to the deadzone when the posture of UAV changes drastically. In that case, the ground receiver is unable to receive part of the signal, and the channel coefficient will experience a specific fade duration. This phenomenon is defined as PVF in this study.
\textcolor{red}{
On this basis, the relative PVF coefficient can be modeled by the roll, pitch, and yaw components as
}
\begin{equation}
\small
\color{red}{
{C^{\rm{P}}}\left( t \right){\rm{ }} = {\rm{ }}{C_\omega }\left( {\omega (t),{\theta _{{\rm{H}},\omega }}} \right){C_\gamma }\left( {\gamma (t),{\theta _{{\rm{H}},\gamma }}} \right){C_\varphi }\left( {\varphi (t),{\theta _{{\rm{H}},\varphi }}} \right)
}
\end{equation}
\textcolor{red}{
and three components denoted by ${C_\vartheta(t)}(\vartheta,{\theta_{{\rm{H}},\vartheta}})$ can be further expressed as
}
\vspace{-8mm}
\begin{spacing}{2.3}
\begin{equation}
\small
\color{red}{
{C_\vartheta}\left( {\vartheta(t),{\theta_{{\rm{H}},\vartheta}}} \right) = \left\{ {\begin{array}{*{20}{c}}
{1}&&{\vartheta(t)}\in{\left( {0,\frac{\pi-{\theta_{{\rm{H}},\vartheta}}}{2}} \right) \bigcup \left( {\frac{3\pi+{\theta_{{\rm{H}},\vartheta}}}{2},2\pi} \right)}
\\
{\cos \left({\frac{\pi}{2{\theta_{{\rm{H}},\vartheta}}}\vartheta(t)+\frac{\pi \left( \theta_{\rm{H}},\vartheta-\pi \right)}{4{\theta_{{\rm{H}},\vartheta}}}} \right)}&&{\vartheta(t)}\in{\left( \frac {\pi- {\theta_ {\rm{H},\vartheta} } } {2}, \frac{ \pi+{\theta_{\rm{H},\vartheta} } } {2} \right)}
\\
{0}&&{\vartheta(t)}\in{\left( {\frac{{\pi  + {\theta _{{\rm{H}},\vartheta }}}}{2},\frac{{3\pi  - {\theta _{{\rm{H}},\vartheta }}}}{2}} \right)}
\\
{\cos \left({\frac{\pi}{2{\theta_{{\rm{H}},\vartheta}}}\vartheta(t)+\frac{\pi \left( 5{\theta_{{\rm{H}},\vartheta}}-3\pi \right)}{4{\theta_{{\rm{H}},\vartheta }}}} \right)}&&{\vartheta(t)}\in{\left( {\frac{3\pi-{\theta_{{\rm{H}},\vartheta}}}{2},\frac{3\pi+{\theta_{{\rm{H}},\vartheta}}}{2}} \right)}
\end{array}} \right.
}
\label{eq.PVF}
\end{equation}
\end{spacing}
\vspace{-10mm}
\noindent
\textcolor{red}{
where $\theta_{{\rm{H}},\vartheta}$ is the projection of the half-power beam width (HPBW) of UAV antenna on each coordinate axis, and $\vartheta \in \{ \omega,\gamma,\varphi \}$. Thus, the PVF effect is described in 3D to ensure roll, pitch, and yaw angle variations can be captured.}
The coefficient is equal to one at the beginning and end of the periodic function, and the posture angle belongs to the interval $(\pi /2+{\theta_{{\rm{H}},\vartheta}}/2,3\pi / 2-{\theta_{{\rm{H}},\vartheta}}/2)$. If ${\theta_{{\rm{H}},\vartheta}}$ is equal to zero, then the coefficient is zero when the posture angle is in the interval $(\pi /2,3\pi /2)$. If ${\theta_{{\rm{H}},\vartheta}}$ is greater than or equal to $\pi$, then the coefficient is equal to 1 only when $\vartheta$ is in the set $\left\{ 0 \right\}\bigcup \left\{ 2\pi \right\}$, other posture angles will cause occlusion from fuselage.

\subsection{Path Angles and Phases}
The proposed model determines the NLoS AoDs and AoAs in different propagation sectors, i.e., the NUS and the NGS, respectively. The corresponding angles can be calculated through the geometric relationship between Tx and the first-bounce scatterers (FBS), or the last-bounce scatterers (LBS) and Rx, respectively. 
\textcolor{red}{
However, different from the conventional stochastic models such as the twin-cluster model \cite{ZhuQM18TC}, the proposed model calculates the AoDs in a deterministic way based on the RT procedure. The FBSs in the proposed model are assumed to be distributed on the wing, nose, and so on, as shown in Fig.~\ref{scenario}.}
The FBSs are dominantly determined by the fuselage structure rather than the scattering environment, and are independent of LBSs. Therefore, AoDs under the FSE should be obtained through deterministic method rather than stochastic approach. Note that radiation field measurement is a precise method to obtain practical AoDs from fuselage, but complicated and time-consuming. Actually, the ray-level angle parameters, i.e., $\alpha _{n,m}^{\rm{Tx}}\left( t \right)$ and $\beta _{n,m}^{\rm{Tx}}\left( t \right)$ in (\ref{eq.h_LoS}) and (\ref{eq.h_NLoS}), is barely accessible in measurement due to limited spatial angle resolution in the existing multi-antenna system. Thus, the RT technique could be adopted as an alternative method to aid the angle calculation.

In the process of RT, the sub-paths reflected by the scatterers on the fuselage can be tracked, and the FSE directional vector of UAV can be defined as
\begin{equation}
\small
{\bf{D}}_{n,m}(t)={{\mathbf{L}}^{\rm{Tx}}}(t)-{{\bf{L}}_{n,m}}(t)
\end{equation}
where ${{\bf{L}}_{n,m}}(t)$ denotes the location of the scattering point on UAV for the $m$-th sub-path. Then, NLoS AoDs of fuselage-scattered sub-paths and the LoS path angles can be determined deterministically as
\begin{equation}
\small
\alpha _{n,m}^{\rm{Tx}}\left( t \right)={{\arctan }_{2}}\left( \frac{{\boldsymbol{e}}_{y} \cdot {\bf{D}}_{n,m}(t)} {{\boldsymbol{e}_{x}} \cdot {\bf{D}}_{n,m}(t)} \right)
\end{equation}
\begin{equation}
\small
\beta _{n,m}^{\rm{Tx}}\left( t \right)=\arcsin \left( \frac{{\boldsymbol{e}_{z}} \cdot {\bf{D}}_{n,m}(t)}{\| {\bf{D}}_{n,m}(t) \|} \right)
\end{equation}
\begin{equation}
\small
\alpha _{l}^{\rm{Tx}}\left( t \right)={{\arctan }_{2}}\left( \frac{{\boldsymbol{e}_{y}} \cdot {{\mathbf{L}}^{\rm{Rx}}}(t)-{{\boldsymbol{e}_{y}} \cdot {{\mathbf{L}}^{\rm{Tx}}}(t)}} {{\boldsymbol{e}_{x}} \cdot {{\mathbf{L}}^{\rm{Rx}}}(t)-{\boldsymbol{e}}_{x} \cdot {{\mathbf{L}}^{\rm{Tx}}}(t)} \right)	
\end{equation}
\begin{equation}
\small
\beta _{l}^{\rm{Tx}}\left( t \right)=\arcsin \left( \frac{{\boldsymbol{e}_{z}} \cdot {{\mathbf{L}}^{\rm{Rx}}}(t)-{{\boldsymbol{e}_{z}} \cdot {{\mathbf{L}}^{\rm{Tx}}}(t)}} {\sqrt{ \| {{\mathbf{L}}^{\rm{Rx}}}(t)-{{\mathbf{L}}^{\rm{Tx}}}(t) \| }} \right)
\end{equation}
\begin{equation}
\small
\alpha _{l}^{\rm{Rx}}\left( t \right)=\alpha _{l}^{\rm{Tx}}\left( t \right)+\pi 	
\end{equation}
\begin{equation}
\small
\beta _{l}^{\rm{Rx}}\left( t \right)=-\beta _{l}^{\rm{Tx}}\left( t \right)	
\end{equation}
where ${{\arctan }_{2}}\left( \cdot  \right)$ denotes the four-quadrant inverse tangent operation, and ${\boldsymbol{e}}_{x}$, ${\boldsymbol{e}}_{y}$ and ${\boldsymbol{e}}_{z}$ are the coordinate axis base vectors, $\left\| \cdot  \right\|$ denotes the Euclidean distance of a location vector. The proposed model can adequately describe the scattering environment near the fuselage based on the FSE vectors provided by deterministic methods, e.g., the RT method. Besides, the AoDs will also be influenced by UAV velocity and posture, the associated effect will be compensated when computing the phase.

Although AoAs can be obtained by the location vector of LBS using a similar approach, they are usually generated from a stochastic process with a specified PDF. For example, the wrapped Gaussian PDF and the wrapped Laplacian PDF in \cite{3GPP20}, the uniform PDF in \cite{QuaDRiGa21}, and the cosine PDF in \cite{ChengX20IoTJ}. In this paper, the initial AoAs are generated by a uniform PDF.

After angle parameters are generated, the geometric relationship between Tx, Rx, and scatterers is determined, then the phase parameter of each path can be calculated. In the proposed model, the LoS and NLoS phases consist of three parts, i.e.,
\begin{equation}
\small
	\Phi _{pq}^{\rm{LoS}}\left( t \right)=\Phi _{\rm{I},\it{pq}}^{\rm{LoS}}\left( t \right)+\Phi _{\rm{D},\it{pq}}^{\rm{LoS}}\left( t \right)+\Phi _{\rm{A},\it{pq}}^{\rm{LoS}}\left( t \right)	
\end{equation}
\begin{equation}
\small
\Phi_{pq,n,m}^{\rm{NLoS}}\left( t \right) = \Phi_{\rm{I},\it{pq,n,m}}^{\rm{NLoS}}\!\left( t \right)\! +\! \Phi_{\rm{D},\it{pq,n,m}}^{\rm{NLoS}}\!\left( t \right)\!+\!\Phi_{\rm{A},\it{pq,n,m}}^{\rm{NLoS}}\!\left( t \right)\!
\end{equation}
where {\small $\Phi _{{\rm I},pq / {\rm I},pq,n,m}^{\rm{LoS/NLoS}}$} are the initial random phases for each LoS path and NLoS sub-path, {\small$\Phi _{{\rm D},pq / {\rm D},pq,n,m}^{\rm{LoS/NLoS}}\left( t \right)$} are the Doppler phases caused by the distance variation, and {\small $\Phi _{{\rm A},pq / {\rm A},pq,n,m}^{\rm{LoS/NLoS}}\left( t \right)$} are the spatial modified phases for MIMO antennas.
The initial phases of LoS and NLoS component can be expressed as
\begin{equation}
\small
\Phi _{\rm{I},\it{pq}}^{\rm{LoS}}\left( t \right)=\left[
\begin{matrix}
{{e}^{\rm{j}\Phi _{\it{pq},\rm{LoS}}^{\rm{VV}}}} & 0  \\
0 & {{e}^{\rm{j}\Phi_{\it{pq},\rm{LoS}}^{\rm{HH}}}}  \\
\end{matrix} \right]\rm{U}\left( 0,2\pi  \right)	
\end{equation}
\begin{equation}
\small
\Phi _{\rm{I},\it{pq,n,m}}^{\rm{NLoS}}\!\left( t \right)\!=\!\left[\!
\begin{matrix}
{{e}^{\rm{j}\Phi _{\it{pq,n,m}}^{\rm{VV}}}} \!&\! \sqrt{{{\kappa }_{n\!,m}}^{\!-\!1}}{{e}^{\rm{j}\Phi _{\it{pq,n,m}}^{\rm{VH}}}}  \\
\sqrt{{{\kappa }_{n\!,m}}^{\!-\!1}}{{e}^{\rm{j}\Phi _{\it{pq,n,m}}^{\rm{HV}}}}\! &\! {{e}^{\rm{j}\Phi _{\it{pq,n,m}}^{\rm{HH}}}}  \\
\end{matrix} \!\right]\rm{U}\!\left( 0,2\pi \right)
\end{equation}
where ${\Phi_{\it{pq}}^{\rm{VV}}}$, ${\Phi _{\it{pq}}^{\rm{HH}}}$, ${\Phi _{\it{pq}}^{\rm{VH}}}$ and ${\Phi _{\it{pq}}^{\rm{HV}}}$ are initial phases of different polarization combinations, which describe the polarization phase variation for the LoS and NLoS paths, and ${{\kappa }_{n,m}}$ is the cross polarization power ratio. Besides, $\rm{U}\left( 0,2\pi  \right)$ is the random variable obeying the uniform distribution over $\left[ 0,\rm{ }2\pi  \right)$, which describes the initial signal phase. Note that the $\Phi _{\rm{I},\it{pq}}^{\rm{LoS}}\left( t \right)$ and $\Phi _{\rm{I},\it{pq,n,m}}^{\rm{NLoS}}\left( t \right)$ have considered the random polarization phase for each path.

The time-variant Doppler phase is a fundamental variable to consider channel non-stationarity, which usually be expressed as the product of frequency shift and time. However, it is physically inconsistent due to the discrete and stochastic generated angles \cite{Patzold17TWC}. Therefore, this study defines the Doppler phases as the phase change caused by relative distance variation. The corresponding Doppler phase components of LoS and NLoS paths can be obtained by
\begin{equation}
\small
\Phi_{{\rm D},pq}^{\rm{LoS}}\left( t \right) = \frac{2\pi}{\lambda} \left| \left( {{\mathbf{L}}^{\rm{Tx}}}(t)-\mathbf{e}_{p}(t) \right) - \left({\mathbf{L}}^{\rm{Rx}}(t) - \mathbf{e}_{q}(t) \right) \right|
\end{equation}
\begin{equation}
\small
\color{red}{
\Phi_{{\rm D},pq,n,m}^{\rm{NLoS}}(t)\!=\!\frac{2\pi}{\lambda} \left( \!d_{n,m}\!+\! (h-{\xi}_{\rm NGS})/\sin{\beta_{l}^{\rm Tx}}(t) \!+\! \left| {\boldsymbol{b}_{p,n,m}}(t) \right|\!+\!\left| {\boldsymbol{a}_{q,n,m}(t)} \right| \right) }
\end{equation}
Note that the Doppler phase constantly changes with the path length variation caused by the movement of both Tx and Rx. The Doppler effect on each path is reflected by uniformly pairing the transceivers in the calculation.

After considering the non-stationarity caused by distance variation, the rotation of the MIMO antenna array should be regarded as due to not only the 3D arbitrary velocity but also the fuselage posture change. The modified antenna phase terms for LoS and NLoS paths are given as
\begin{equation}
\small
\Phi _{\rm{A},\it{pq}}^{\rm{LoS}}\left(t\right) = \frac{2\pi}{\lambda}\left({\boldsymbol{r}}_{p}^{\rm{Tx}} {{{\bf{R}}}^{\rm{P}}}(t) {\boldsymbol{s}}_{l}^{\rm{Tx}}(t) + {\boldsymbol{r}}_{q}^{\rm{Rx}} {{{\bf{R}}}^{\rm{Rx}}}(t) {\boldsymbol{s}}_{l}^{\rm{Rx}}(t) \right)
\end{equation}
\begin{equation}
\small
\Phi _{\rm{A},\it{pq,n,m}}^{\rm{NLoS}}\left(t\right) = \frac{2\pi}{\lambda}\left({\boldsymbol{r}}_{p}^{\rm{Tx}} {{{\bf{R}}}^{\rm{P}}}(t) {\boldsymbol{s}}_{n,m}^{\rm{Tx}}(t)\!+\!{\boldsymbol{r}}_{q}^{\rm{Rx}} {{{\bf{R}}}^{\rm{Rx}}}(t) {\boldsymbol{s}}_{n,m}^{\rm{Rx}}(t) \right)
\end{equation}
where ${\boldsymbol{r}}_{p}^{\rm{Tx}}$ and ${\boldsymbol{r}}_{q}^{\rm{Rx}}$ are the position vectors of the $p$-th transmitting antenna and the $q$-th receiving antenna relative to the center of the antenna array, respectively, 
\textcolor{red}{
the velocity rotation matrix ${{\bf{R}}}^{\rm{Rx}}(t)$ aims to correct the position vector under different movement direction and keep consistent with the original definition \cite{ZhuQM19MAP}. Note that even if the velocity is zero, the UAV can have a posture change, which also causes a rotation of the coordinate system. Therefore, a specific posture matrix ${{\bf{R}}^{\rm{P}}}(t)$ is introduced to describe the posture of UAV. The definition of roll, yaw, and pitch angles can be found in Fig.~\ref{scenario}. The transfer matrix from the coordinate system of Tx to that of Rx is shown as}
\begin{equation}
\small
	\begin{aligned}
  & {{\left[ x\ y\ z \right]}^{\rm{T}}} = {{{\bf{R}}}^{\rm{P}}}(t){{\left[ \tilde{x}\ \tilde{y}\ \tilde{z} \right]}^{\rm{T}}} = {{{\bf{R}}}_{z}}{\bf{R}}_{y}{\bf{R}}_{x}{ \left[ \tilde{x}\ \tilde{y}\ \tilde{z} \right] ^{\rm{T}} } \\
 & =\left[ \begin{matrix}
   \cos \omega\cos \varphi   & \cos \omega\sin \varphi  \sin \gamma -\sin \omega\cos \gamma  & \cos \omega\sin \varphi  \cos \gamma +\sin \omega\sin \gamma   \\
   \sin \omega\cos \varphi   & \sin \omega\sin \varphi  \sin \gamma +\cos \omega\cos \gamma  & \sin \omega\sin \varphi  \cos \gamma -\cos \omega\sin \gamma   \\
   -\sin \varphi   & \cos \varphi  \sin \gamma  & \cos \varphi  \cos \gamma   \\
\end{matrix} \right]\left[ \begin{matrix}
   {\tilde{x}}  \\
   {\tilde{y}}  \\
   {\tilde{z}}  \\
\end{matrix} \right] \\
\end{aligned}
\label{eq.xyz*R^P}
\end{equation}
\vspace{-4mm}

Through changing the roll, yaw, and pitch angles, i.e., $\omega $, $\varphi $, and $\gamma $ in the proposed posture matrix, the posture variation of UAV can be incorporated into the channel model. Thus, it can describe the posture variation of UAV and modify the spatial phase variation correctly.
\section{Statistical Properties Analysis}
\subsection{ACF}
The channel fading sensitivity with respect to the time difference can be estimated from the fluctuation of temporal ACFs. ACF can be calculated by the summation of LoS and NLoS components as
\begin{equation}
\small
{{R}_{pq}}\left( t;\Delta t \right) = {R}_{pq}^{\rm{LoS}}\left( t;\!\Delta t \right) + {R}_{pq,n}^{\rm{NLoS}}\left( t;\Delta t \right)
\label{eq.ACF}
\end{equation}
where the LoS component of ACF ${R}_{pq}^{\rm{LoS}}\left( t;\!\Delta t \right)$ and the NLoS component ${R}_{pq,n}^{\rm{NLoS}}\left( t;\Delta t \right)$ are assumed independent with each other, and can be expressed as
\begin{equation}
\small
{R}_{pq}^{\rm{LoS}}\left( t;\Delta t \right) = {h^{\rm LoS}_{pq}(t)}^{*} h^{\rm LoS}_{pq}(t+\Delta t){e}^{ {\rm j} \left(\Phi_{pq}^{\rm LoS}( t+\Delta t)-\Phi_{pq}^{\rm LoS}(t)\right)}
\end{equation}
\begin{equation}
\small
{R}_{pq,n}^{\rm NLoS}\left( t;\Delta t \right) \!=\! {\rm E}\!\left[\sum\limits_{m=1}^{M}{{h^{\rm NLoS}_{pq,n,m}(t)}^{*}h^{\rm NLoS}_{pq,n,m}(t+\Delta t) {e}^{{\rm j}\left( \!\Phi_{pq,n,m}^{\rm NLoS}(t+\Delta t)-\Phi_{pq,n,m}^{\rm NLoS}(t) \right) }}\right]
\label{eq.ACF_NLoS}
\end{equation}
where ${\rm E}[\cdot]$ denotes the expectation operator.
\subsection{PDP}
\textcolor{red}{
The time-variant PDP describes the power distribution under the delay of multi-path components, and it can be expressed by \cite{Patzold11} }
\begin{equation}
\small
\color{red}{
{{S}_{pq}}\left(t, \tau \right)=\sum\limits_{n=1}^{N(t)}\sum\limits_{m=1}^{M}{{{\left| h_{pq,n,m}^{\rm NLoS}(t) \right|}^{2}}\delta \left( \tau -\tau _{n,m}^{\rm{NLoS}}(t) \right)}. }
\end{equation}
\textcolor{red}{
Note that arbitrary trajectories and velocities of Tx, as well as the fuselage attitude, will have an effect on the trend of the PDP.}
\vspace{-2mm}
\subsection{LCR and AFD}
The LCR is the expected rate of a stochastic process crosses a given signal $r$ with a positive or negative slope within one second, which can be used to evaluate the communication quality of the channel. The LCR of a channel can be calculated by
\begin{equation}
\small
N_{\zeta}\left( r,t \right) = \int_{0}^{\rm{+}\infty }{{\dot{r}}}{p_{\zeta \dot{\zeta}}}(r,\dot{r},t) d\dot{r}	
\end{equation}
where $r$ is a given level greater than 0, ${p}_{\zeta \dot{\zeta}}(r,\dot{r},t)$ denotes the joint PDF of the time-variant channel fading envelope $\zeta (t)$ and its time derivative $\dot{\zeta}(t)$.

\textcolor{red}{
The AFD is the expected value of the length of the time duration in which a stochastic process is below a given signal level $r$, and can be calculated by means of \cite{Patzold11} }
\begin{equation}
\small
\color{red}{
T_{\zeta}\left( r,t \right) = \frac{\int_{0}^{r}{p_{\zeta}\left(x\right)dx}} {N_{\zeta}\left(r,t\right)}  }	
\end{equation}
\textcolor{red}{
where $p_{\zeta}(x,t)$ is the PDF of the time-variant channel fading envelope.}
\vspace{-2mm}
\subsection{SI}
For non-stationary channels, the SI needs to be defined, which indicates the maximum time duration over which the wide-sense-stationary condition is valid. SI can be obtained by counting the maximum duration ${{T}_{s}}$ of the channel output at different time snapshots ${{t}_{a}}$. When the correlation coefficient of two average PDPs is equal to $\varepsilon_0 $, which is generally set as 80\%, the stationary interval can be calculated from the equation $\varepsilon \left( {{t}_{a}},{{T}_{s}} \right)=\varepsilon_0$, where the correlation coefficient of two averaged PDPs can be further expressed as
\begin{equation}
\small
\varepsilon \left( {{t}_{a}},{{T}_{s}} \right)\!=\!\frac{\int{{{{\bar{\Lambda }}}_{pq}}\left( {{t}_{a}},\tau  \right){{{\bar{\Lambda }}}_{pq}}\left( {{t}_{a}}\!+\!{{T}_{s}},\tau  \right)\rm{d}\tau }}{\max \left\{ \int{{{{\bar{\Lambda }}}_{p\!q}}^{2}\left( {{t}_{a}},\tau  \right)\rm{d}\tau },\!\int{{{{\bar{\Lambda }}}_{p\!q}}^{2}\left( {{t}_{a}}\!+\!{{T}_{s}},\!\tau  \right)\rm{d}\tau } \right\}}
\end{equation}
where the averaged PDPs can be calculated by
\begin{equation}
\small
{{\bar{\Lambda }}_{pq}}\left( {{t}_{a}},\tau  \right)=\frac{1}{{{N}_{a}}}\sum\limits_{a}^{a+{{N}_{a}}-1}{{{\left| {{h}_{pq}}\left( {{t}_{a}},\tau  \right) \right|}^{2}}}	
\end{equation}
\textcolor{red}{
where ${{t}_{a}}$ denotes the time of the $a$-th drop, a drop is the minimum time interval limited by the computational precision, and ${{N}_{a}}$ denotes the total number of averaged PDPs. The averaged PDPs over successive drops of time are sequentially compared to determine the SI.}

\section{Numerical Results and Discussions}
In this section, the statistical properties of the proposed U2G model are analyzed through numerical simulations. We study the effects of FSE, UAV posture, and PVF on U2G channel characteristics. In addition, the practicability of the model is verified by comparing the measured data with the simulation results.

\begin{figure}
\centering
\includegraphics[height=6cm]{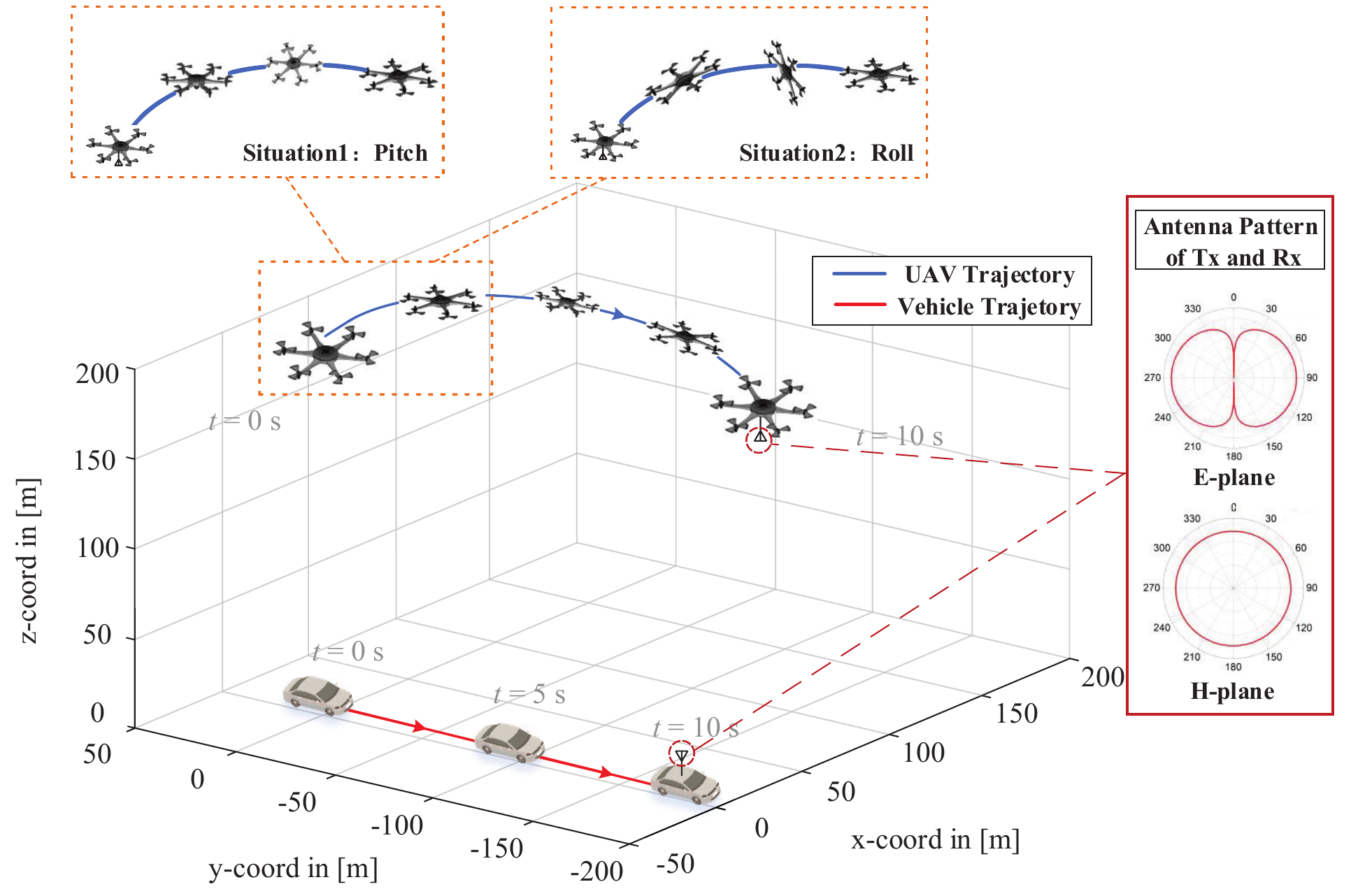}
\caption{
\color{red}{
Antenna pattern and trajectories of two mobile terminals.}
}
\label{trajectory}
\vspace{-6mm}
\end{figure}

\textcolor{red}{
In the simulation, the hexacopter UAV and the ground vehicle are equipped with the typical dipole antenna, and the antenna pattern is shown in Fig.~\ref{trajectory}.} 
The carrier frequency is 2.4~GHz, the height of the UAV is 150~m, and the UAV moves in a curve around a semicircle with a radius of 100~m at a speed of 30~m/s. The ground vehicle moves in a straight line at a constant velocity of 20~m/s.
\textcolor{red}{
The trajectories of the two terminals are also shown in Fig.~\ref{trajectory}.Based  on the massive RT simulation results on the hexacopter UAV, the AAoD and EAoD of the first NLoS path are set to $\pi /3$ and $\pi /12$, respectively.}
In addition, we simulated the changes of the pitch angle and roll angle during the flight of the UAV. The rotation trajectories are also shown in the figure, where the posture rotation speed is set to $\dot{\gamma} =\pi/4 \text{ rad/s}$ in the Simulation~1 and $\dot{\omega} =\pi/4 \text{ rad/s}$ in the Simulation~2, respectively.

\begin{figure}
\centering
\includegraphics[height=6cm]{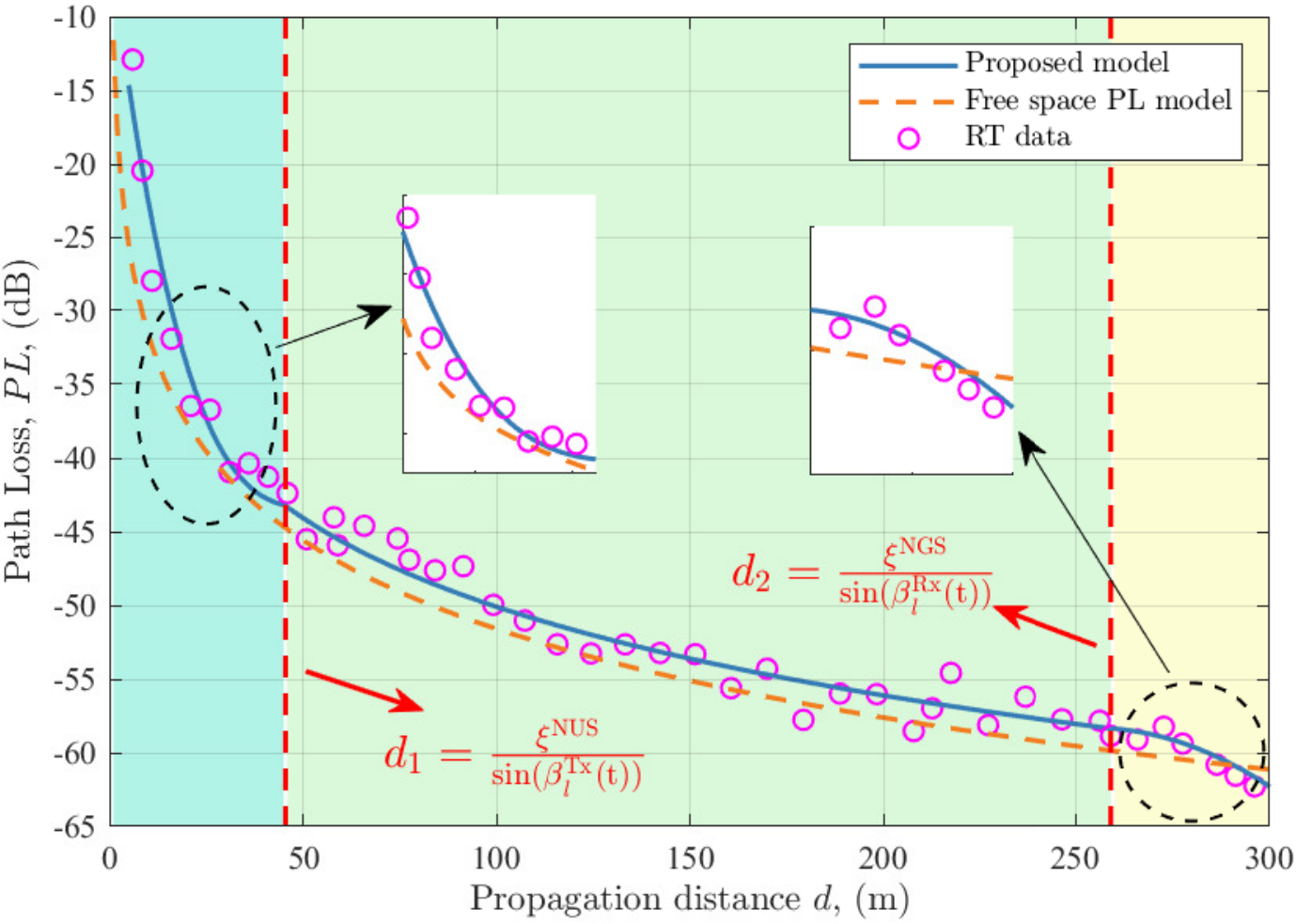}
\vspace{-2mm}
\caption{
\color{red}{PLs of proposed model, free space model, and RT method.}}\label{PL}
\vspace{-3mm}
\end{figure}

\textcolor{red}{
Taking the first NLoS path as an example, the proposed PL model and the free-space model are compared in Fig.~\ref{PL}. PLs for the NUS, FSL, and NGS are identified by different colors, and the height-related segment points $d_1$ and $d_2$ are also marked. When ${{h^{{\rm{Tx}}}} - {\xi ^{{\rm{NUS}}}} \le h}$, the UAV fuselage makes a part of propagation signals reflected, resulting in a slightly higher receiving power compared with the free-space model. When ${{\xi ^{{\rm{NGS}}}} \le h \le {h^{{\rm{Tx}}}} - {\xi ^{{\rm{NUS}}}}}$, the trend of proposed PL is consistent with the free-space model. When ${h \le {\xi ^{{\rm{NGS}}}}}$, the near-ground scattering effect is considered in a stochastic way and it results in a rapid attenuation of receiving power.}
Additionally, we use the RT method to obtain the PL of the UAV channel in the same scenario. The simulated PL has a good agreement with the RT data, which proves the practicability of the proposed model.

\begin{figure}
\centering
\subfigure[]{
\begin{minipage}{7cm}
	\centering
	 \includegraphics[scale=0.53]{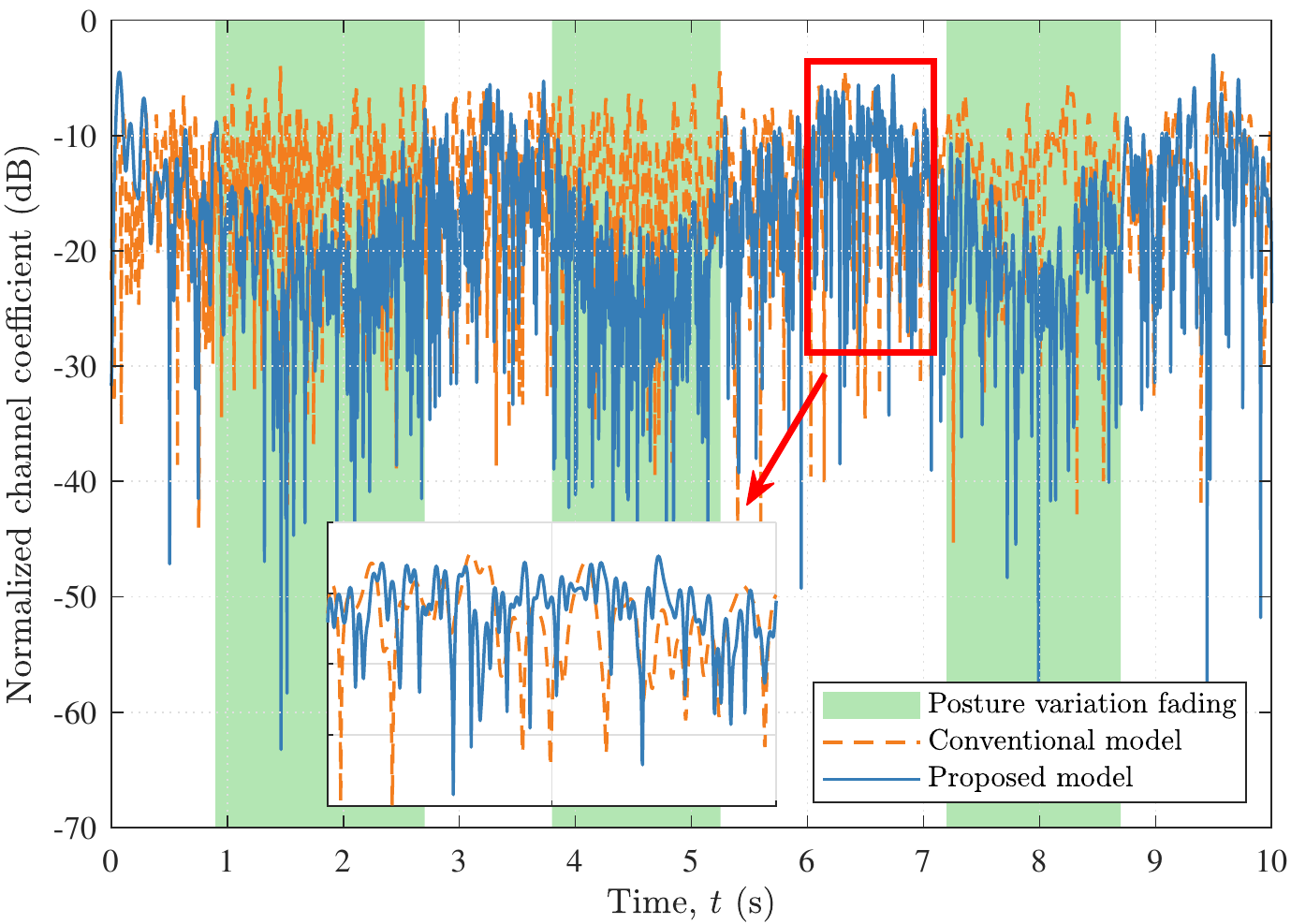}
	\end{minipage}
}
\subfigure[]{
\begin{minipage}{7cm}
	\centering
	 \includegraphics[scale=0.53]{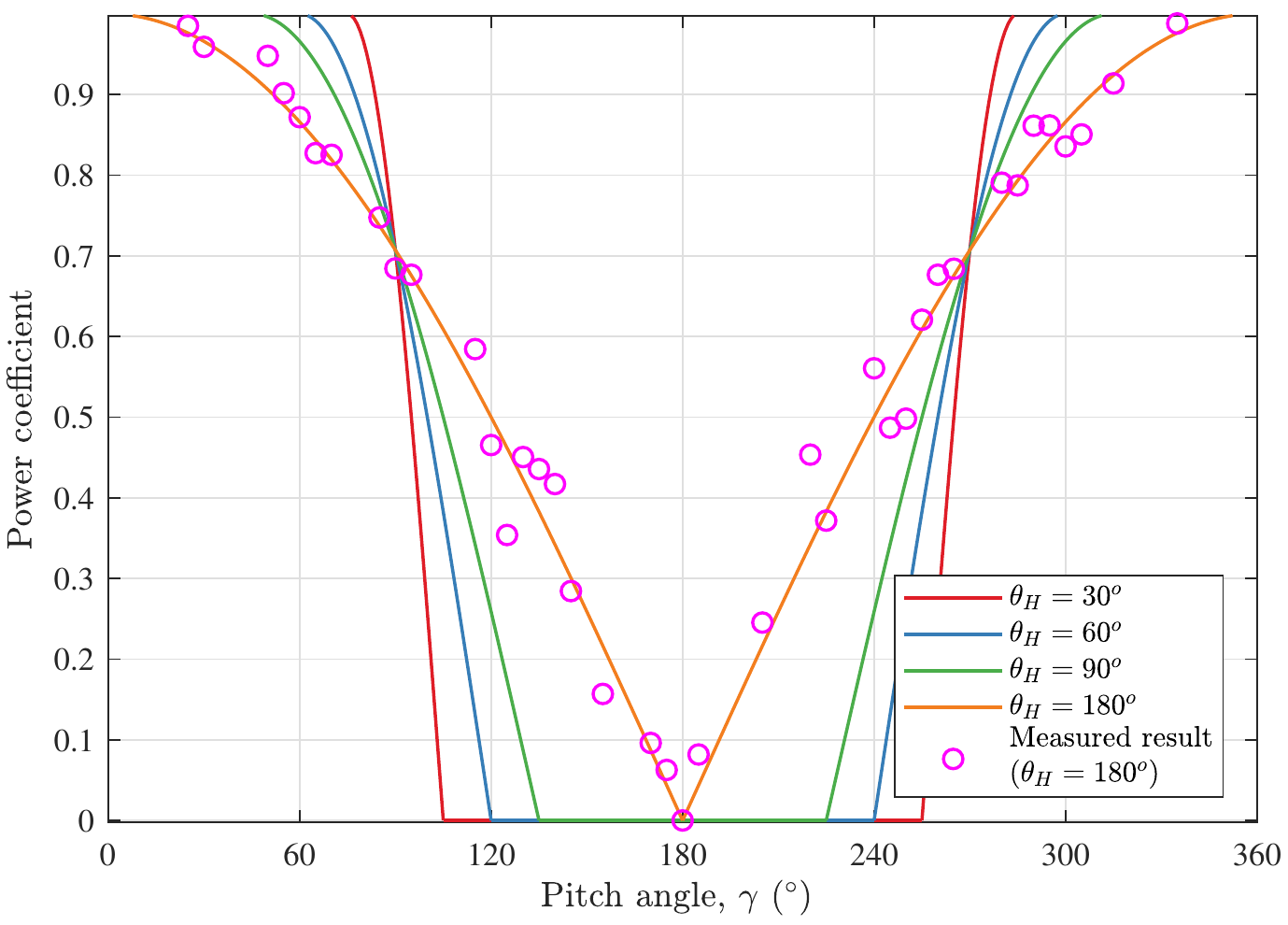}
	\end{minipage}
}
\vspace{-4mm}
\caption{(a) Channel coefficients and (b) PVF coefficient of proposed model.}
\label{Channel coefficients and PVF}
\vspace{-6mm}
\end{figure}

The normalized channel coefficients $h(t)$ of the proposed model and the conventional model in \cite{ZhuQM19MAP} are shown in Fig.~\ref{Channel coefficients and PVF}(a).
\textcolor{red}{
We choose the pitch angle $\gamma$ as an example to illustrate the impact of UAV posture variation. Compared with the model in \cite{ZhuQM19MAP}, the channel coefficient of the proposed model exhibits periodic fluctuations. In the periods of 0.9~s to 2.7~s, 3.8~s to 5.2~s, and 7.2~s to 8.7~s, the channel coefficient firstly decreases and then increases. Fig.~\ref{Channel coefficients and PVF}(b) shows the power coefficient variation under different postures. As the pitch angle increases, the antenna radiation is gradually blocked by the fuselage, and the signal power decreases. Different HPBWs determine the range of posture angle in which the PVF phenomenon occurs. The wider the HPBW, the narrower the signal zero areas. In order to verify the correctness of the PVF coefficient proposed in this paper, we measured the power of the UAV equipped with the dipole antenna at different pitch angles $\gamma$. The power coefficients are normalized for comparison. The measured data are marked with red circles in the figure, which agree with the analysis results calculated by (\ref{eq.PVF}).
}

\begin{figure}
\centering
\includegraphics[height=5.9cm]{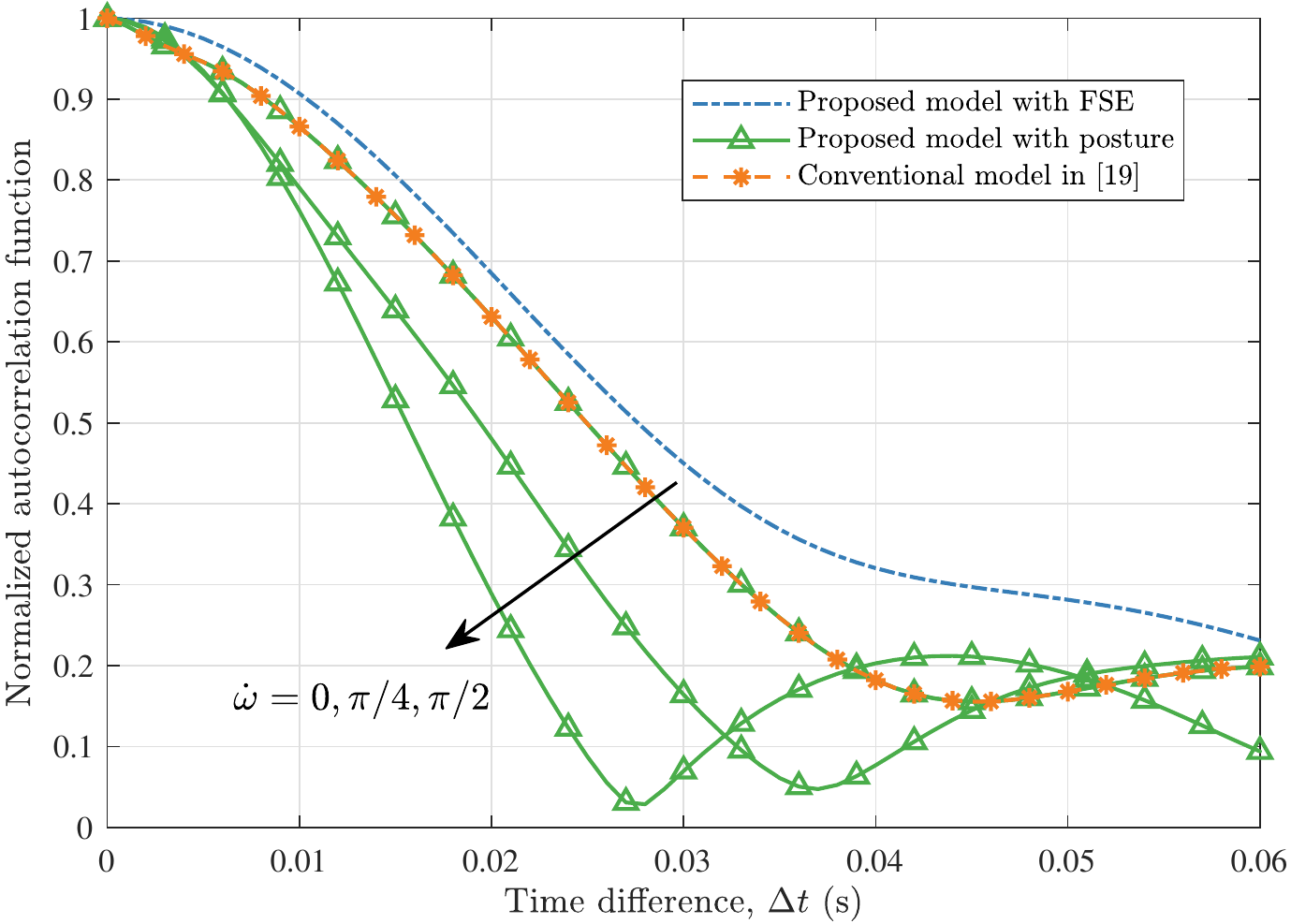}
\caption{ACF with proposed model and conventional model.}
\vspace{-1mm}
\label{ACF with proposed model and conventional model.}
\vspace{-6mm}
\end{figure}

In order to verify the influence of UAV posture variation and FSE on channel characteristics, the temporal ACFs of the proposed model and the conventional model in \cite{ZhuQM19MAP} are provided in Fig.~\ref{ACF with proposed model and conventional model.}. The analytical results can be calculated by (\ref{eq.ACF})-(\ref{eq.ACF_NLoS}). The decline rate of ACF with FSE is slower than that without FSE because FSE incorporates more deterministic information into the modeling and reduces the randomness. Besides, different roll angles result in various trends of ACFs. The pattern change of ACF has a regular relationship with the variation of roll angle. The ACF of the proposed model approaches the result of the model in \cite{ZhuQM19MAP} when the posture variation becomes smaller. The simulated ACFs have a good consistency with the corresponding analytical results, which proves the correctness of the proposed channel model and ACF derivations. In addition, the same simulations are conducted with pitch variation and the conclusion is similar to that of roll angle.

\begin{figure}
\centering
\includegraphics[height=5.9cm]{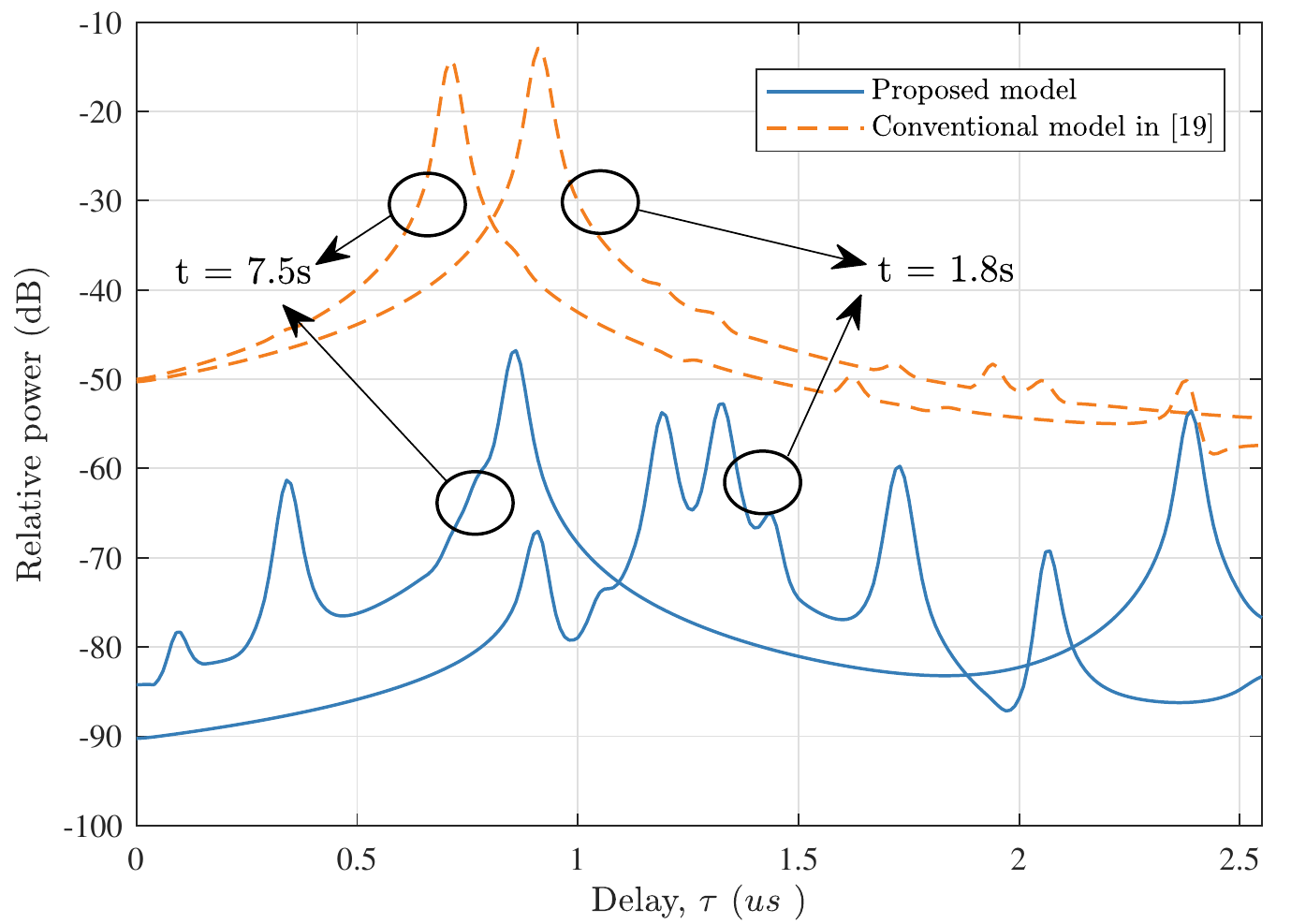}
\caption{PDP of proposed model and conventional model.}
\vspace{-3mm}
\label{PDP}
\end{figure}

Fig.~\ref{PDP} shows the normalized PDPs of the proposed model and the conventional model at different moments. In the conventional model, the strongest receiving power can be determined to be the LoS component, and the power decreases with the evolution of time delay. It is worth noting that 1.8~s and 7.5~s are the moments when the transmitting antenna of UAV rotates away from the receiving terminals. At these moments, the LoS component of proposed model is occluded by fuselage seriously. Thus, the receiving power is significantly attenuated, and the LoS and NLoS components cannot be distinguished in Fig.~\ref{PDP}. It is proved that the UAV posture variation has a significant impact on the channel, which helps guide the design of the U2G communication system.

\begin{figure}
\centering
\subfigure[]{
\begin{minipage}{7cm}
	\centering
	 \includegraphics[scale=0.6]{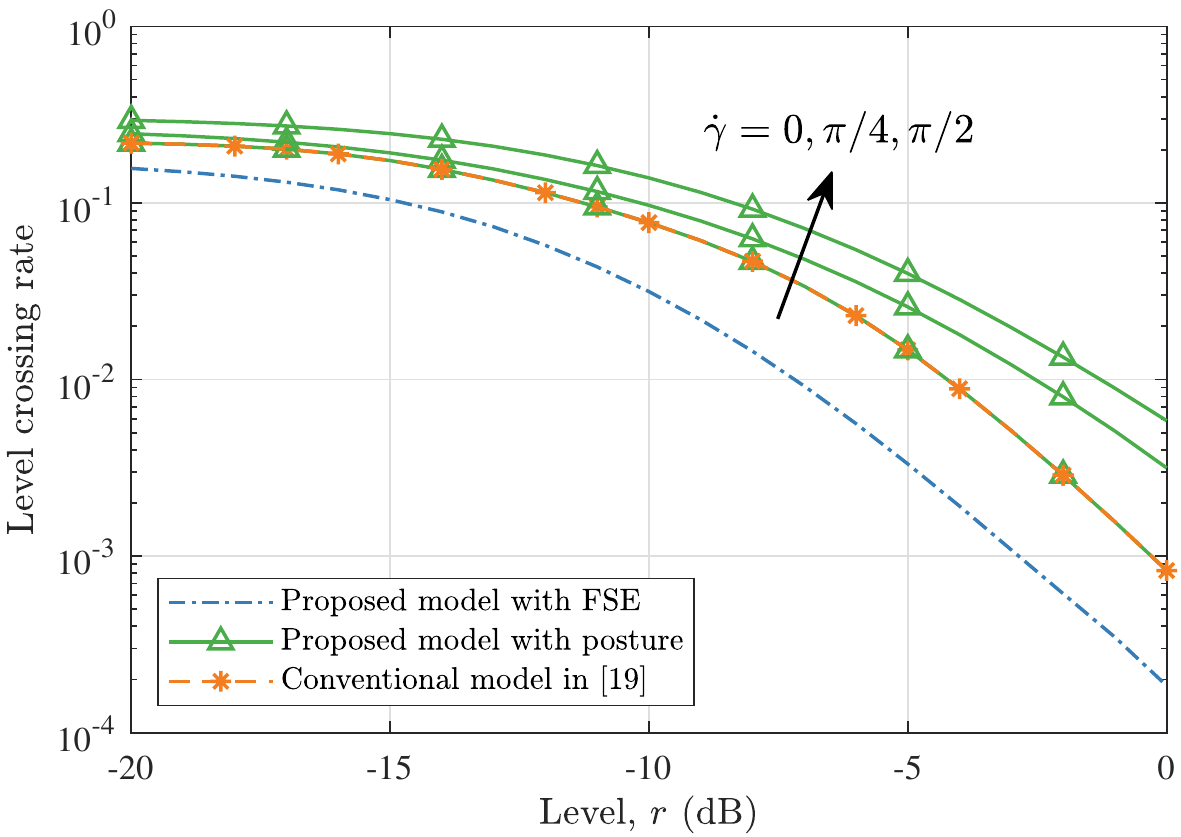}
	\end{minipage}
}
\subfigure[]{
\begin{minipage}{7cm}
	\centering
	 \includegraphics[scale=0.6]{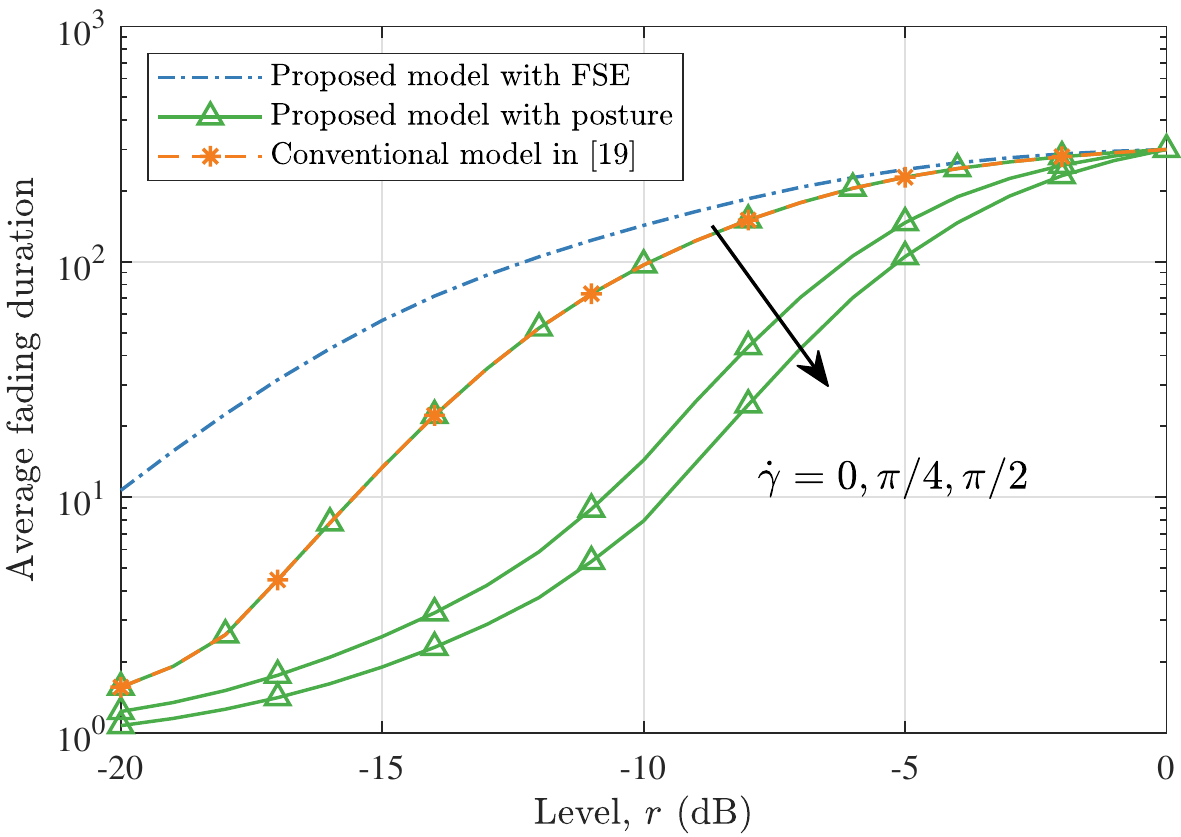}
	\end{minipage}
}
\vspace{-4mm}
\caption{(a) LCR and (b) AFD of proposed model and conventional model.}
\label{LCR and AFD of proposed model and conventional model.}
\vspace{-6mm}
\end{figure}

In order to explain the time fluctuation of UAV channel envelope, LCR and AFD of the proposed model and conventional model are shown in Fig.~\ref{LCR and AFD of proposed model and conventional model.}. The fading envelope tends to be smaller than the level most of the time when the threshold level increases. The rate at which the fading envelope breaks through the level decreases, while the average duration of the fading envelope below the level increases. Thus, the LCR decreases while the AFD increases as the threshold level increases. The results in Fig.~\ref{LCR and AFD of proposed model and conventional model.}(a) show that the LCR of the proposed model is higher than that of the conventional model when the fuselage posture changes. That is because the UAV posture rotation increases the propagation complexity and makes the channel fading more serious. On the other hand, the LCR of the proposed model considering FSE is lower than that of the conventional model. The possible reason is that the scattering points are fixed on the fuselage, making the channel fade relatively gentle. We can get the same conclusion from AFD in Fig.~\ref{LCR and AFD of proposed model and conventional model.}(b).

\begin{figure}
\centering
\subfigure[]{
\begin{minipage}{7cm}
	\centering
	 \includegraphics[scale=0.6]{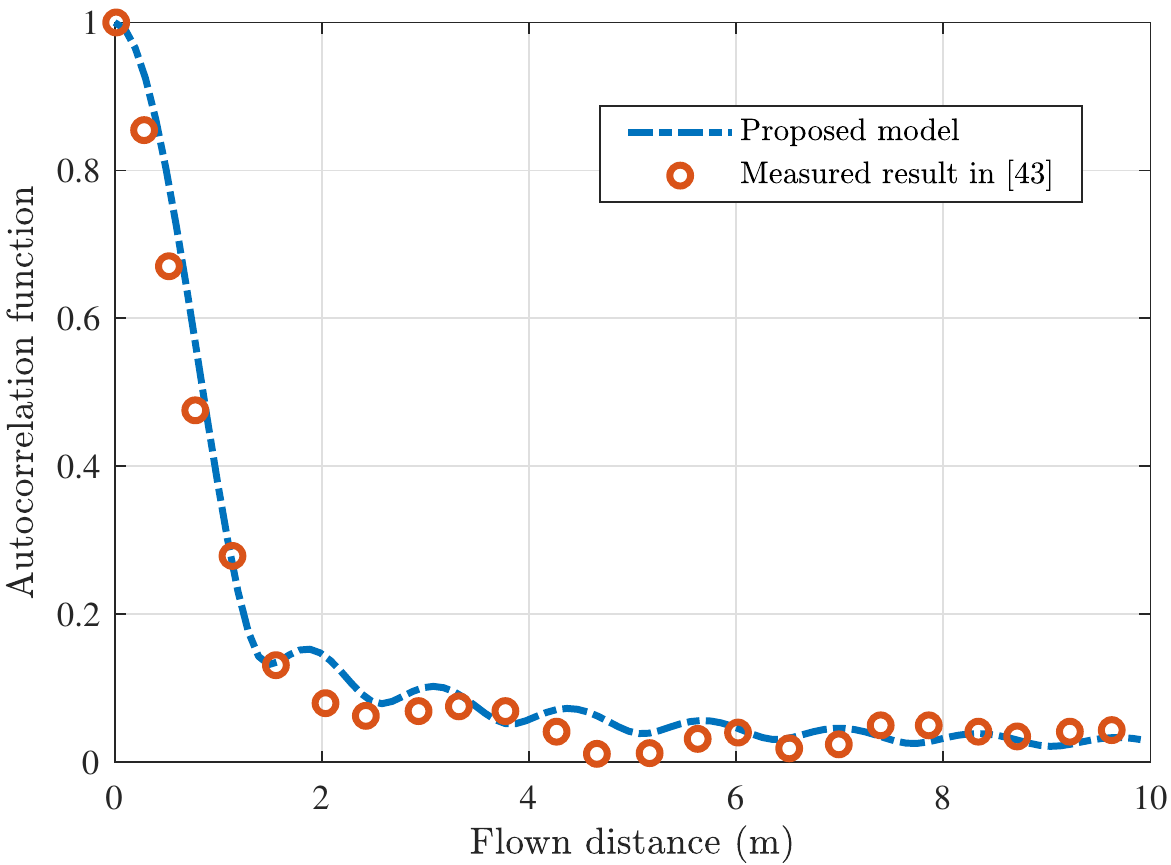}
	\end{minipage}
}
\subfigure[]{
\begin{minipage}{7cm}
	\centering
	 \includegraphics[scale=0.6]{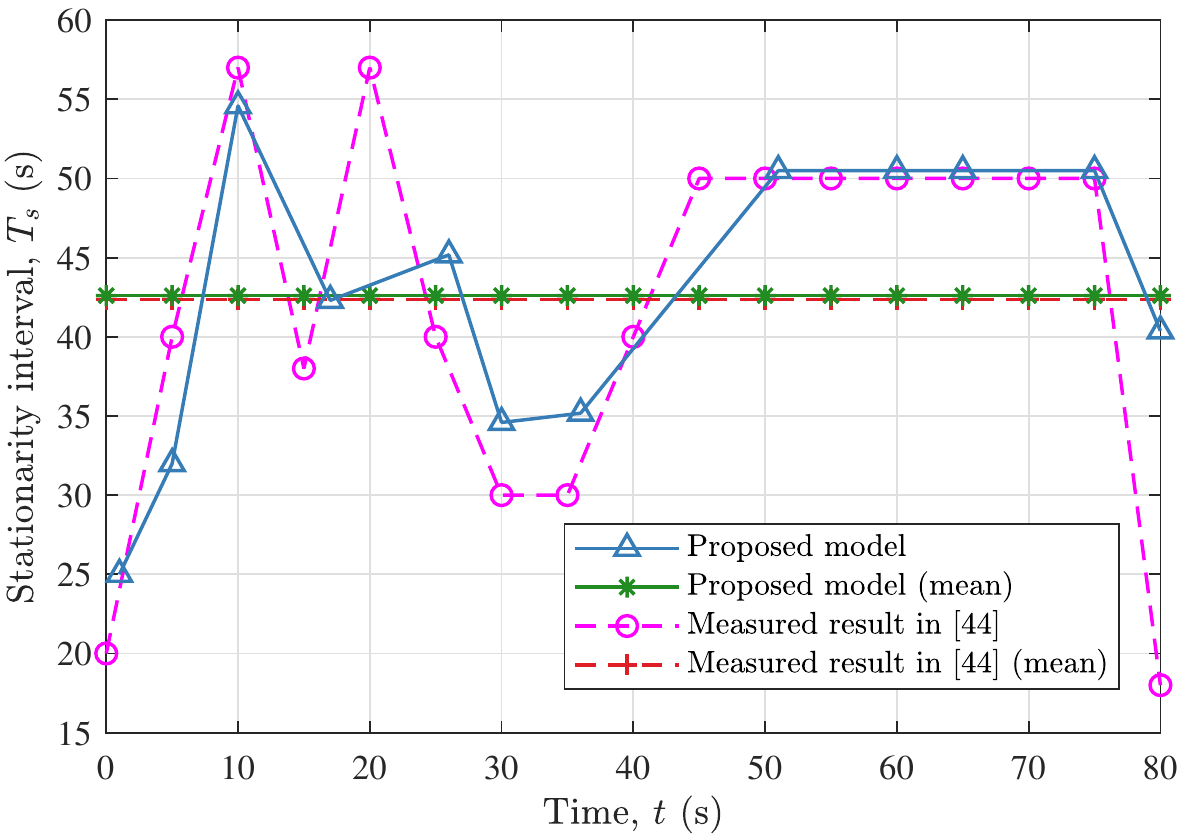}
	\end{minipage}
}
\vspace{-4mm}
\caption{(a) ACF and (b) SI of proposed model and other measured results.}
\vspace{-2mm}
\label{ACF and SI of proposed model and measured results in other reference.}
\end{figure}

To illustrate the compatibility of proposed model, we change the channel parameters to the scenario of the measurement campaign in \cite{Simunek13TAP} and \cite{Kanatas17TAP}, respectively. The comparison of ACFs is shown in Fig.~\ref{ACF and SI of proposed model and measured results in other reference.}(a), where the measured result is obtained with variant flown distances, and the simulated result is obtained from the proposed model. The carrier frequency is 2.5~GHz, and the speed of UAV and vehicle are 40~m/s and 10~m/s, respectively. It can be found that the ACF of proposed model matches well with the measurement data. Fig.~\ref{ACF and SI of proposed model and measured results in other reference.}(b) shows the SI of the proposed model and measured results, where the simulated and measured SI are close, and the mean SI is highly consistent. Therefore, the proposed model is compatible with simulating U2G channels under different scenarios.

\begin{figure}
\centering
\subfigure[]{
\begin{minipage}{7cm}
	\centering
	 \includegraphics[scale=0.55]{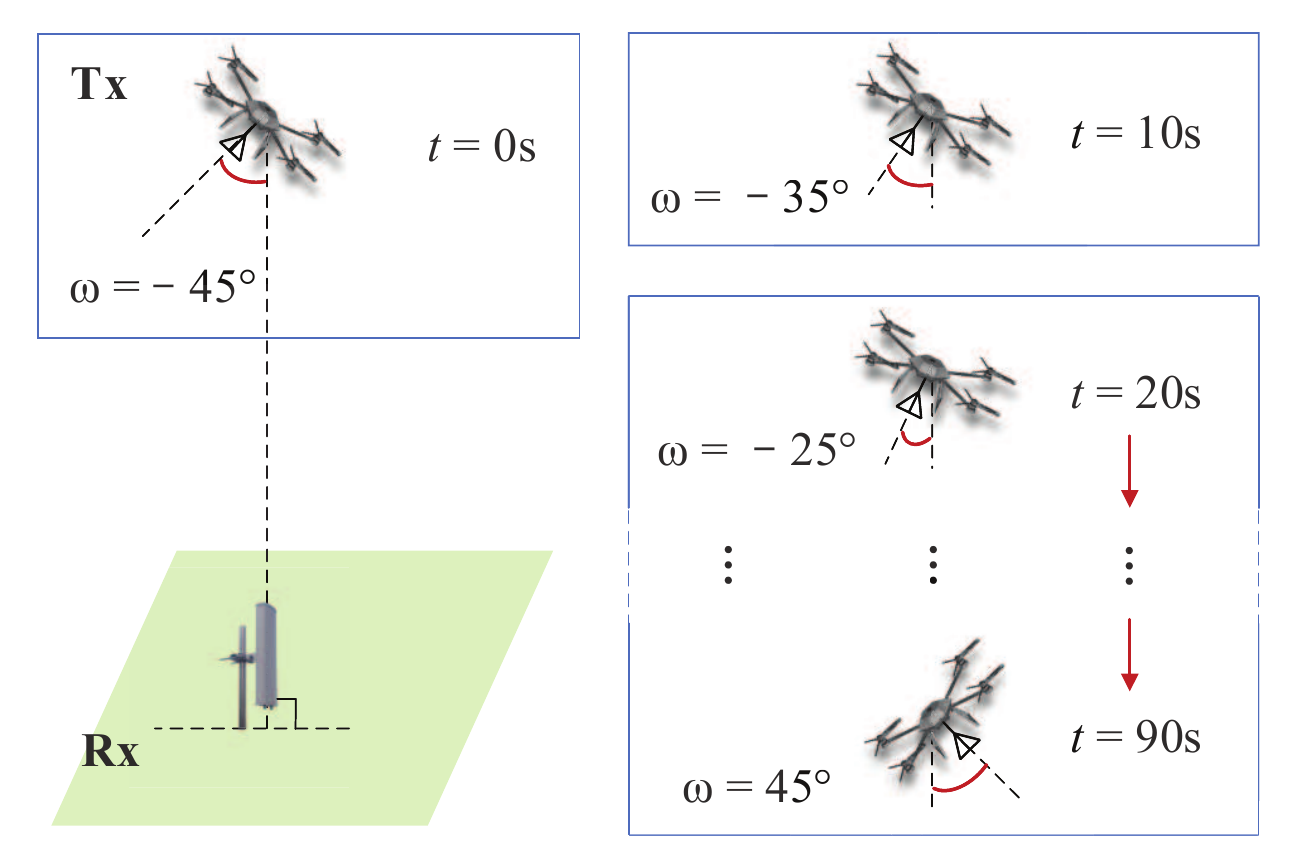}
	\end{minipage}
}
\subfigure[]{
\begin{minipage}{7cm}
	\centering
	 \includegraphics[scale=0.5]{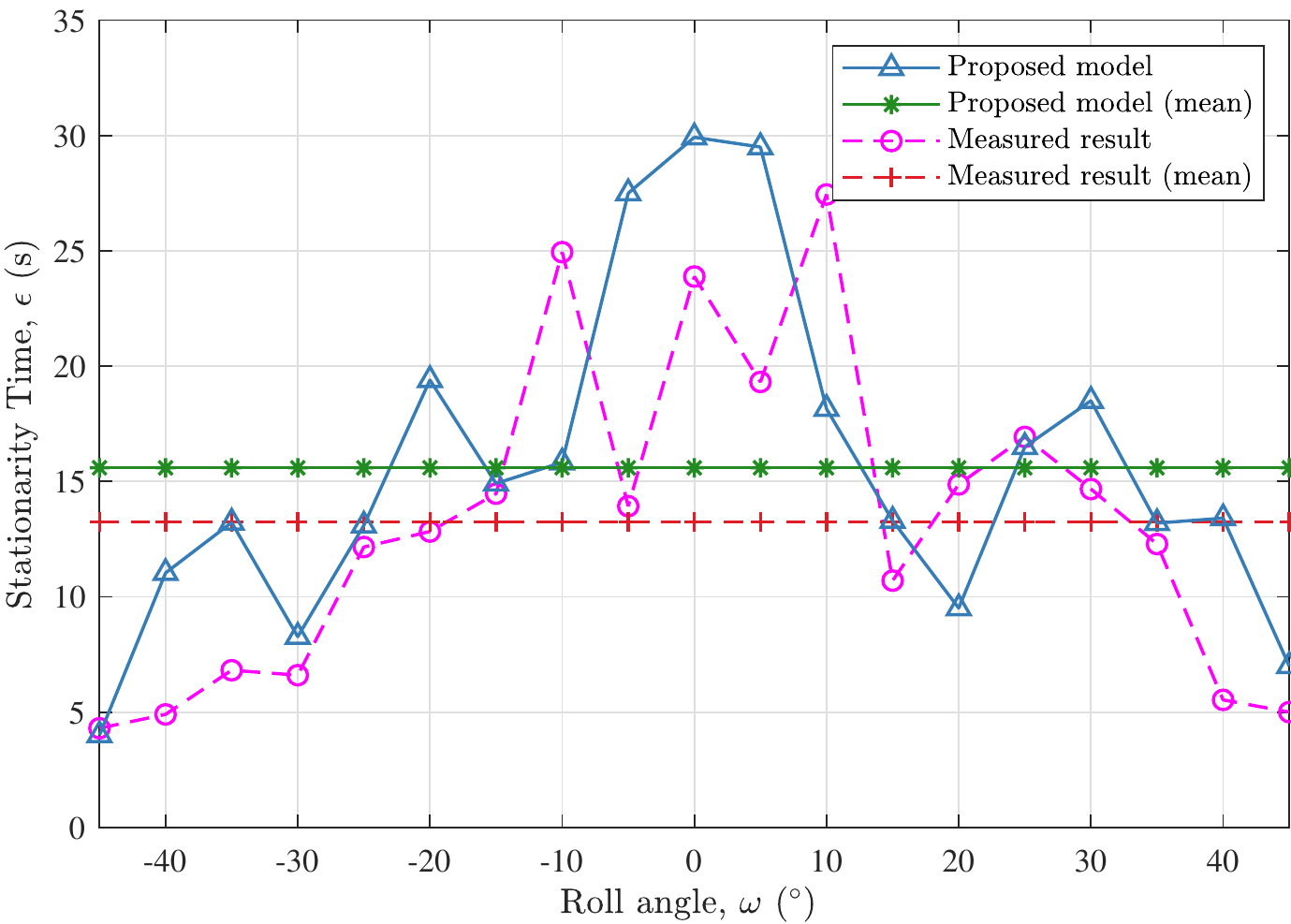}
	\end{minipage}
}
\vspace{-3mm}
\caption{Proposed U2G (a) measurement setup and \color{red}{(b) measured results}.}
\vspace{-6mm}
\label{Proposed U2G measurement and results.}
\end{figure}

In order to further verify the non-stationary characteristics caused by UAV posture variation in the proposed model, we conduct a measurement campaign to capture the SI with different
\textcolor{red}{roll angles $\omega$,} as shown in Fig.~\ref{Proposed U2G measurement and results.}(a) and (b). Note that the SI in Fig.~\ref{Proposed U2G measurement and results.}(b) reaches the larger value when the roll angle is between -10$^{\circ}$ to 10$^{\circ}$, and the larger roll angle leads to the smaller SI. 
Both simulated and measured SI varies with the roll angle, which indicates that the SI is closely related to posture variation. The simulation results are in good agreement with the measured ones, which shows that the proposed model can reflect the non-stationary characteristics of U2G channel caused by posture variations.

\section{Conclusion}
This paper has proposed a non-stationary U2G MIMO channel model by thoroughly considering the impact of UAV posture variation and FSE. By applying the hybrid parameter generation approach, the generation procedure of segmented time-variant channel parameters have been given. The PL has shown different trends from conventional models in the NUS and NGS. The UAV posture variation has been considered by transforming the coordinate system and introducing the posture matrix. Meanwhile, the impact of PVF on the channel parameters has been modeled by a segmented function. Based on these improvements, the time-evolution continuity of the channel parameters has been achieved and the realistic U2G channel can be characterized by the proposed model.
Besides, the critical statistical properties have been analyzed and compared with conventional channel models. Posture variation and FSE have been proven to significantly impact the ACF, PDP, LCR, and AFD. Furthermore, the practicability of the proposed model is validated by the PVF coefficient and SI from the corresponding measurement campaign. The proposed model can be applied to diverse U2G communication scenarios by adjusting parameters, i.e., UAV to vehicle, UAV to pedestrian, and UAV to base station. Thus, it may have the potential value for designing, optimizing, and evaluating UAV MIMO communication systems.


\newpage
\section{Biography Section}

\vspace{-66pt}
\begin{IEEEbiography}[{\includegraphics[width=0.9in,height=1.25in,clip,keepaspectratio]{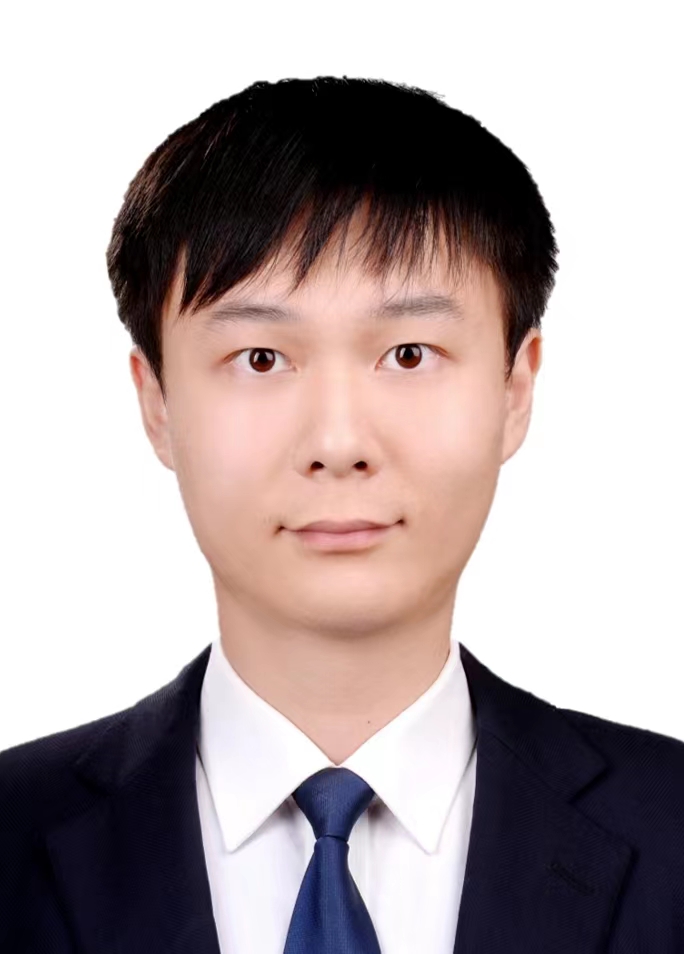}}]{Boyu Hua}
received the B.S. degree in physics from Nanjing Normal University, China, in 2014, and the M.S. degree in electronic communication engineering from Nanjing University of Aeronautics and Astronautics (NUAA), China, in 2018. Since 2018, he has been an experimentalist and is currently pursuing Ph.D. degree in communication and information systems in NUAA. He received two Best Paper Awards from IEEE IWCMC 2020 and IEEE CSPS 2021. His research interests include wireless channel modeling for B5G and 6G communication.
\end{IEEEbiography}

\vspace{-33pt}
\begin{IEEEbiography}[{\includegraphics[width=0.9in,height=1.25in,clip,keepaspectratio]{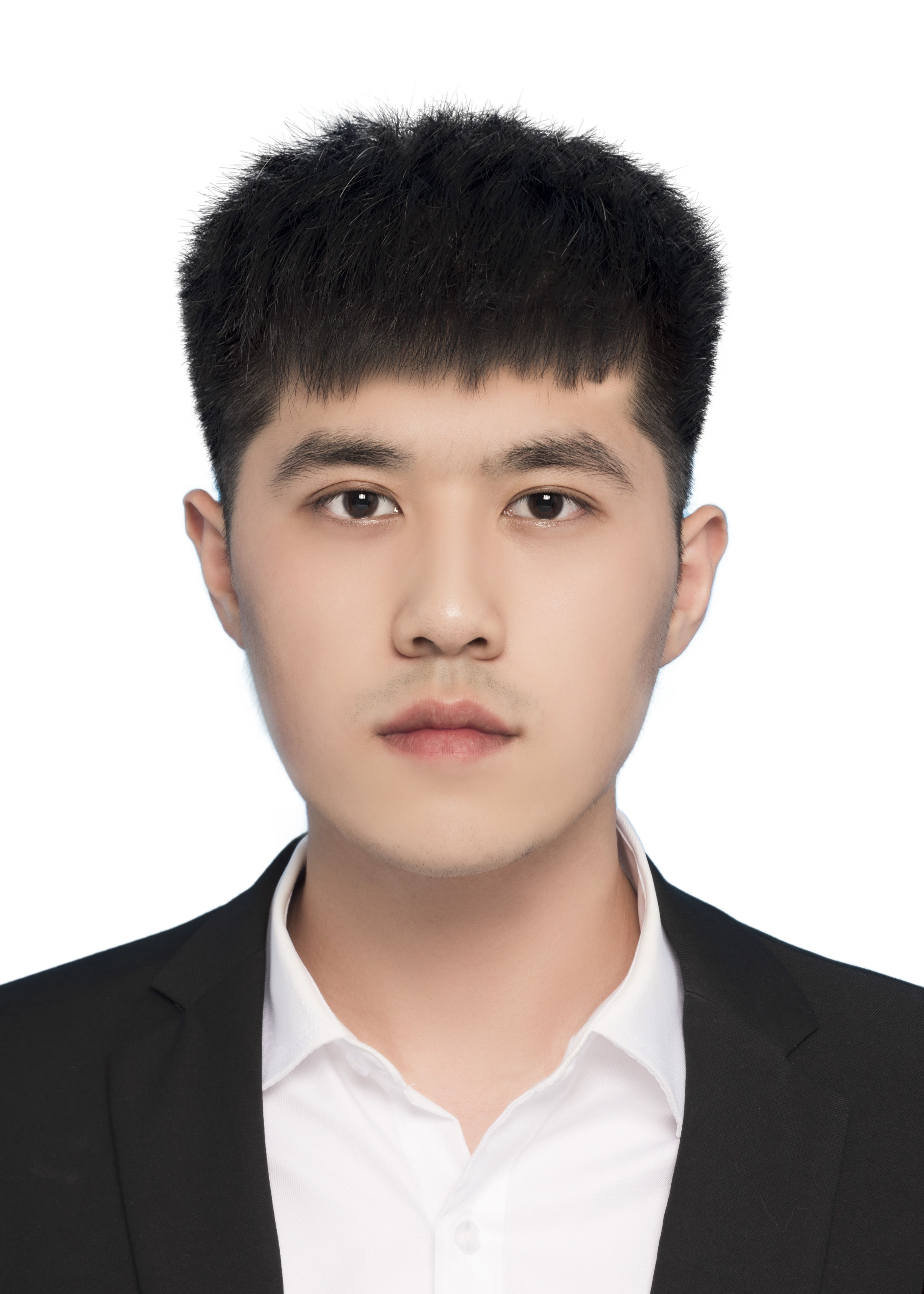}}]{Haoran Ni}
received the B.S. in information engineering from Nanjing University of Information Science \& Technology, Nanjing, China, in 2021. He is currently pursuing the M.S.degree in information and communication engineering in Nanjing University of Aeronautics and Astronautics (NUAA). His research interests include channel modeling for UAV communication systems.
\end{IEEEbiography}

\vspace{-33pt}
\begin{IEEEbiography}[{\includegraphics[width=1in,height=1.25in,clip,keepaspectratio]{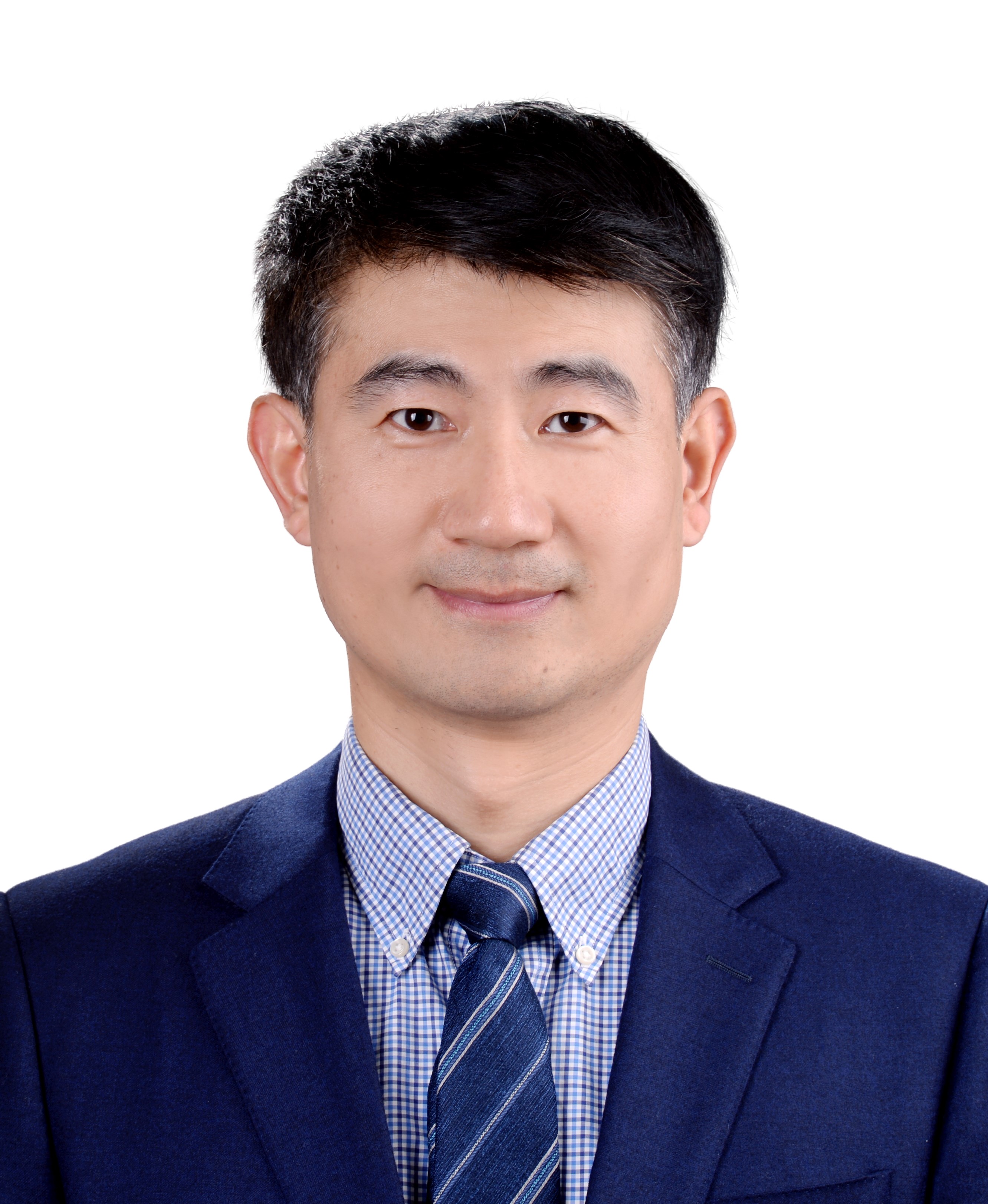}}]{Qiuming Zhu}
received his BS in electronic engineering from Nanjing University of Aeronautics and Astronautics (NUAA), Nanjing, China, in 2002 and his MS and PhD in communication and information system from NUAA in 2005 and 2012, respectively. Since 2021, he has been a professor in the Department of Electronic Information Engineering, NUAA. From Oct.~2016 to Oct.~2017, June~2018 to Aug.~2018 and June~2018 to Aug.~2018, he was also an academic visitor at Heriot Watt University, Edinburgh, U. K. He has authored or coauthored more than 120 articles in refereed journals and conference proceedings and holds over 40 patents. His current research interests include channel sounding, modeling, and emulation for the fifth/sixth generation (5G/6G) mobile communication, vehicle-to-vehicle (V2V) communication and unmanned aerial vehicles (UAV) communication systems.
\end{IEEEbiography}

\vspace{-33pt}
\begin{IEEEbiography}[{\includegraphics[width=0.9in,height=1.25in,clip,keepaspectratio]{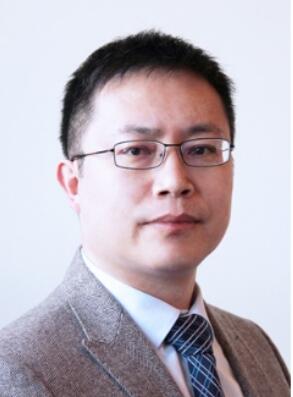}}]{Cheng-Xiang Wang}
received the B.Sc. and M.Eng. degrees in communication and information systems from Shandong University, Jinan, China, in 1997 and 2000, respectively, and the Ph.D. degree in wireless communications from Aalborg University, Aalborg, Denmark, in 2004. He has been with Heriot-Watt University, Edinburgh, U.K., since 2005, where he was promoted to a Professor in 2011. In 2018, he joined Southeast University, Nanjing, China, as a Professor. He is currently a part-time Professor with Purple Mountain Laboratories, Nanjing. He has authored four books, three book chapters, and more than 430 papers in refereed journals and conference proceedings, including 25 highly cited papers. He has also delivered 22 invited keynote speeches/talks and nine tutorials in international conferences. His current
research interests include wireless channel measurements and modeling, 6G wireless communication networks, and applying artificial intelligence to wireless communication networks.
\end{IEEEbiography}

\vspace{-33pt}
\begin{IEEEbiography}[{\includegraphics[width=1.1in,height=1.25in,clip,keepaspectratio]{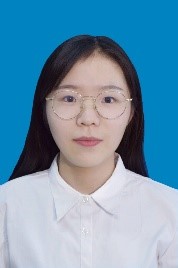}}]{Tongtong Zhou}
received the B.S. in comminication engineering from Nanjing University of Information Science \& Technology, Nanjing, China, in 2020. She is currently pursuing the M.S. degree in information and communication engineering in Nanjing University of Aeronautics and Astronautics (NUAA). Her research interests include channel modeling for UAV communication systems.
\end{IEEEbiography}

\vspace{-33pt}
\begin{IEEEbiography}[{\includegraphics[width=1in,height=1.25in,clip,keepaspectratio]{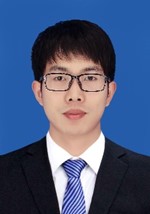}}]{Kai Mao}
received the B.S. in information engineering and M.S. degree in electronics and communication engineering from the Nanjing University of Aeronautics and Astronautics (NUAA), Nanjing, China, in 2016 and 2019, respectively. He is currently pursuing the Ph.D. degree in communication and information systems. His research interests include channel sounding, modeling for UAV communication systems and wireless channel emulators.
\end{IEEEbiography}

\vspace{-33pt}
\begin{IEEEbiography}[{\includegraphics[width=0.9in,height=1.25in,clip,keepaspectratio]{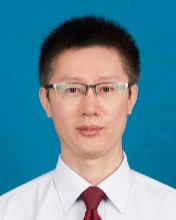}}]{Junwei Bao}
received his master and Ph.D. degrees from Sichuan University and Nanjing University of Aeronautics and Astronautics (NUAA) in 2005 and 2019, respectively. He is currently a Lecturer in College of Physics, NUAA. His current research interests include wireless communication theory and wireless channel modeling.
\end{IEEEbiography}

\vspace{-33pt}
\begin{IEEEbiography}[{\includegraphics[width=1in,height=1.25in,clip,keepaspectratio]{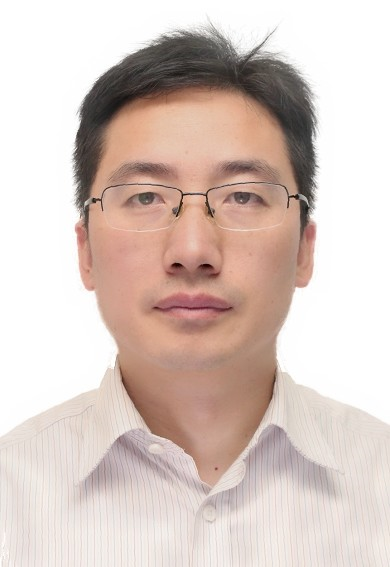}}]{Xiaofei Zhang}
received M.S degree from Wuhan University, Wuhan, China, in 2001. He received Ph.D. degrees in communication and information systems from Nanjing University of Aeronautics and Astronautics in 2005. Now, he is a Professor in Electronic Engineering Department, Nanjing University of aeronautics and astronautics, Nanjing, China. His research is focused on array signal processing and communication signal processing. He serves as Editors of the International Journal of Digital Content Technology and its Applications, International Journal of Technology and Applied Science, Journal of Communications and Information Sciences, Scientific Journal of Microelectronics, and International Journal of Information Engineering.
\end{IEEEbiography}
%
%
%
%
%
%

\end{document}